\documentclass[12pt]{iopart}
\usepackage{iopams}
\usepackage{epsfig}

\newcommand\bra[1]{\left\langle #1 \right |}
\newcommand\ket[1]{\left|#1\right\rangle}
\newcommand\braket[2]{\left\langle #1 \right.\left|#2\right\rangle}
\newcommand\me[3]{\left\langle #1 \right | #2 \left|#3\right\rangle}
\newcommand\fracalph{{\tilde \alpha}}
\newcommand\Avec{{\bi A}}
\newcommand\Fvec{{\bi F}}
\newcommand\Pvec{{\bi P}}

\newcommand\Vvec{{\bi V}}
\newcommand\Wvec{{\bi W}}
\newcommand\rvec{{\bi r}}
\newcommand\kvec{{\bi k}}

\newcommand\uvec{{\bi u}}
\newcommand\vvec{{\bi v}}
\newcommand\svec{{\bi s}}
\newcommand\Rvec{{\bi R}}
\newcommand\xvec{{\bi x}}
\newcommand\yvec{{\bi y}}
\newcommand\zvec{{\bi z}}
\newcommand\zerovec{{\bf 0}}
\newcommand\xivec{{\bxi}}

\newcommand\psiin{\psi_{\rm in}}
\newcommand\Rhat{\,\hat{\Rvec}}
\newcommand\ehat{\,\hat{\bi e}}
\newcommand\nablavec{{\bi \nabla}}
\newcommand\Vcal{{\cal V}}
\newcommand\Fcal{{\cal F}}
\newcommand\Wcal{{\cal W}}
\newcommand\Ical{{\cal I}}
\newcommand\ftilde{{\tilde f}}
\newcommand\rhotilde{{\tilde \rho}}
\newcommand\Icalpm{{\cal I}_{\pm}}
\newcommand\Iexpect{{\left\langle\Icalpm\right\rangle}}
\newcommand\Udag{{U^{\dag}}}
\newcommand\sgn{\,{\hbox{\rm sgn}}}

\newcommand\eqref[1]{(\ref{#1})}
\newcommand\re{{\hbox{\rm Re}\,}}
\newcommand\im{{\hbox{\rm Im}\,}}
\newcommand\phihat{\,{\hat{\bphi}}}
\newcommand\xhat{\,\hat{\xvec}}
\newcommand\yhat{\,\hat{\yvec}}
\newcommand\zhat{\,\hat{\zvec}}
\newcommand\khat{\,\hat{\kvec}}
\newcommand\Ptrans{P_{\rm trans}}
\begin{document}
\jl{1} \title{Force and impulse from an Aharonov-Bohm flux
  line}[Aharonov-Bohm force and impulse]

\author{JP Keating \& JM Robbins}

\address{BRIMS, Hewlett-Packard Laboratories, Filton Road, Stoke
  Gifford, Bristol BS34 8QZ, UK\\
  and\\
  School of Mathematics, University of Bristol, University Walk,
  Bristol BS8 1TW, UK}

\begin{abstract} 
  We calculate the force operator for a charged particle in the field
  of an Aharonov-Bohm flux line.  Formally this is the Lorentz force,
  with the magnetic field operator modified to include quantum
  corrections due to anomolous commutation relations.  For stationary
  states, the magnitude of the force is proportional to the product of
  the wavenumber $k$ with the amplitudes of the `pinioned' components,
  the two angular momentum components whose azimuthal quantum numbers
  are closest to the flux parameter $\alpha$.  The direction of the
  force depends on the relative phase of the pinioned components.  For
  paraxial beams, the transverse component of our expression gives an
  exact version of Shelankov's formula [Shelankov A 1998 {\it
    Europhys.~Lett.} {\bf 43}, 623 -- 8], while the longitudinal
  component gives the force along the beam.
  
  Nonstationary states are treated by integrating the force operator
  in time to obtain the impulse operator. Expectation values of the
  impulse are calculated for two kinds of wavepackets.   For slow
  wavepackets, which spread faster than they move, 
  the impulse is inversely proportional to the distance from the flux
  line.  For fast wavepackets, which spread only negligibly before
  their closest approach to the flux line, the impulse is proportional
  to the probability density transverse to the incident direction
  evaluated at the flux line.  In this case, the transverse component
  of the impulse gives a wavepacket analogue of Shelankov's formula.
  The direction of the impulse for both kinds of wavepackets is flux dependent.

  We give two derivations of the force and impulse operators, the first
  a simple derivation based on formal arguments, and the second a
  rigorous calculation of wavepacket expectation values.  We
  also show that the same expressions for the force and impulse are
  obtained if the flux line is enclosed in an impenetrable cylinder,
  or distributed uniformly over a flux cylinder, in the limit that
  the radius of the cylinder goes to zero.

\end{abstract}
%\pacs{00.00, 20.00, 42.10}
\maketitle

\section{Introduction}\label{sec:introduction}
There have been a number of investigations of the force exerted on a
charged particle by an Aharonov-Bohm flux line. Classically, of
course, there is no force, so it, like the Aharonov-Bohm effect
itself, is essentially quantum mechanical, vanishing as
$\hbar\rightarrow 0$.  Olariu \& Popescu (1983, 1985) show that for
certain localized wavepackets (these are the fast wavepackets
described in \Sref{sec:fast-wave-packets} below), the force, along
with the momentum it imparts, is negligible unless the centre of the
wavepacket hits the flux line. Nielson and Hedeg\aa rd (1995) and
Shelankov (2000) compute matrix elements of the force operator for
stationary states of the same energy. Shelankov (1998) calculates the
transverse force on a stationary incident beam of finite angular width
using a paraxial approximation, a result we refer to as {\it
  Shelankov's formula}. This use of the paraxial approximation has
been justified by Berry (1999), who computes the asymptotic deflection
of the beam from an exact representation.  Peshkin (1981, 1989)
computes the expectation value of the force when the flux line is
enclosed in an impenetrable cylinder. Recent interest in this problem
has been stimulated by the analogy with the Iordanskii force
(Iordanskii 1966) exerted on phonons by a vortex in a superfluid (see,
eg, Sonin 1975, Sonin 1997, Thouless \etal 1997, Stone 1999), which
has been the subject of some recent debate.

In this paper we add to these investigations in several ways.  First,
we obtain exact expressions for the Lorentz force operator due
to an Aharonov-Bohm flux line and its matrix elements between
arbitrary stationary states; the restriction to on-shell matrix
elements, between stationary states of the same energy, agrees with
previous results.  We also obtain an exact expression for the time
integral of the force, the impulse operator, and compute its
expectation values for various kinds of wavepackets. We show that the
force operator can be simply derived, formally at least, from purely
kinematic considerations, and give a mathematically rigorous
demonstration to justify the results obtained in this way.

The paper is organized as follows.  In \Sref{sec:force-operator}, we
give a formal derivation of the (vector) Lorentz force operator due to
an Aharonov-Bohm flux line.  Pointing out that the nominal magnetic
field operator, $(\alpha hc/e) \delta^2(\rvec)$, is incompatible with
gauge invariance, we show that that a modification of the canonical
commutation relations restores gauge invariance and yields the Lorentz
force.  An explicit formula for the expectation value for stationary
states follows directly.  On-shell matrix elements between stationary states of
the same energy are seen to coincide with previous results.  The
transverse component of the force is shown to coincide with
Shelankov's formula in the paraxial limit (see also Shelankov (2000)).
We then compute (\Sref{sec:impulse-operator}) the impulse operator,
the integral of the force operator in infinite forward and backward
time, in the position representation.  In \Sref{sec:Iexpectation} we
compute the leading-order expectation value of the impulse for two
kinds of wavepackets. For slow wavepackets (\Sref{sec:slow}), which
spread more quickly than they move, the impulse is inversely
proportional to the initial distance between the wavepacket and the
flux line, with magnitude and direction depending periodically on the
flux parameter $\alpha$.  By treating the deflection of a slow
wavepacket as a classical scattering, we obtain an expression for the
scattering cross-section which, surprisingly, coincides with the exact
result. Fast wavepackets (\Sref{sec:fast-wave-packets}) move more
quickly than they spread, so much so that the fractional increase in
their width as they pass the flux line is small.  In this case, 
the impulse is proportional to the transverse
probability density at the flux line, and its transverse component
gives a wavepacket analogue of Shelankov's formula.

In the Appendix, we give a rigorous derivation of the force and
impulse expectation values for a class of well-behaved wavefunctions.
In \Sref{sec:encl-distr-flux}, we compute matrix elements of the force
operator for two standard regularizations of the Aharonov-Bohm flux
line, the first where the flux is enclosed in an impenentrable
cylinder, and the second where it is distributed uniformly in a
cylindrical tube.  The Aharonov-Bohm results of
\Sref{sec:force-operator} and~\Sref{sec:impulse-operator} are
recovered in the limit that the radius of the cylinder goes to zero.

%As we were completing this paper, we obtained a preprint by Shelankov
%(2000) which contains some of the results in \Sref{sec:force-operator}
%below, in particular the exact expression (\ref{eq:stationary}) for
%the force expectation value for a stationary state, and a derivation
%of the paraxial formula from the transverse component of this
%expression.  Shelankov's calculation, based on the momentum-flow
%tensor, is different to ours.  He also gives some interesting physical
%discussion.

\section{Force operator}\label{sec:force-operator}

Consider a particle of charge $e$ and mass $M$ moving in the
$xy$-plane in a magnetic field along $\zhat$. Quantum mechanically,
the particle is described by the Hamiltonian $H = \frac12
M\Vvec\cdot\Vvec$, where $M\Vvec = \Pvec - e\Avec/c$ is the kinetic
momentum and $\Avec(\rvec)$ is the vector potential. The force, ie the
rate of change of the kinetic momentum, is given by the appropriately
symmetrized Lorentz force operator,
\begin{equation}\label{eq:1}
\Fvec = \frac{1}{\rmi\hbar}[M\Vvec, H] = 
\frac{e}{2c}\left( \Vvec\wedge (B\zhat) -  (B\zhat)\wedge \Vvec\right),
\end{equation}
where the magnetic field operator is defined by
\begin{equation}
  \label{eq:2}
B = \frac{\Phi_0}{2\pi\rmi \hbar^2}[MV_x,MV_y].
\end{equation}
Here $\Phi_0 = 2\pi\hbar c/e$ is the magnetic flux quantum.  The usual
commutation relations for position and canonical momentum lead to
the usual relation between the
magnetic field and the vector potential, namely $B\zhat = \nablavec\wedge \Avec$.

However, this relation is incorrect for the vector potential
of an Aharonov-Bohm flux line (since we are restricting to the plane,
we should perhaps say `flux point', but we will follow conventional usage).
For a flux line at the origin of strength $\alpha \Phi_0$,
and in a circularly
symmetric gauge, the vector potential is given by
\begin{equation}
  \label{eq:16}
  \Avec(\rvec) = \alpha\frac{ \Phi_0}{2\pi r}\phihat.
\end{equation}
As is well known, physically meaningful quantities depend only on the
fractional part, $\fracalph$, of the flux parameter, $\alpha$; a unit
shift in the flux parameter, $\alpha\rightarrow \alpha + 1$, is
equivalent to the gauge transformation $\psi\rightarrow U\psi$, where
\begin{equation}
\label{eq:5}
(U\psi)(r,\phi) = \rme^{\rmi\phi}\psi(r,\phi).
\end{equation}
As a consequence, a physically observable operator $O(\alpha)$ which
depends on the flux parameter must transform under $U$ according to
\begin{equation}
\label{eq:6}
U O(\alpha)\Udag = O(\alpha + 1).
\end{equation}
For example, the kinetic momentum $M\Vvec$ satisfies this relation, as
is easily verified.  As the magnetic field operator (\ref{eq:2}) is
expressed in terms of the commutator of components of kinetic
momentum, it must also satisfy (\ref{eq:6}).  However,
$\nablavec\wedge \Avec$, which is given by
$\alpha\Phi_0\delta^2(\rvec)\zhat$, clearly does not satisfy
(\ref{eq:6}); it is invariant under the gauge transformation $U$ (like
any operator which depends only on position), but is not periodic in
$\alpha$.  It follows that substituting $\nabla\wedge\Avec$ for
the magnetic field in (\ref{eq:1}) cannot give the correct expression for the
Lorentz force operator.

The problem is caused, of course, by the singularity at $r=0$, and can
be avoided by an explicit calculation of the time rate of change of
the expectation value of the kinetic momentum for suitably chosen
wavefunctions.  This is done in the Appendix.  However, we can obtain
the same result more easily and more directly from formal arguments.
If the usual canonical commutation relations lead to a magnetic field
operator which does not transform correctly under gauge
transformations, then the canonical commutation relations must be
modified by the flux line.  In particular, we show below that,
formally, the components of the canonical momentum, $p_x$ and $p_y$,
do not commute -- equivalently, the partial derivatives $\partial_x$
and $\partial_y$ do not commute -- in the presence of nonzero flux.

First, we note that the partial derivatives $\partial_x$ and
$\partial_y$
certainly commute when applied
to smooth wavefunctions.  That is,
\begin{equation}
  \label{eq:8}
  [\partial_x,\partial_y] \psi = 0
\end{equation}
for $\psi(\rvec)$ smooth.  Applying the gauge transformation $U$
to this relation $m$ times, we get
\begin{equation}
  \label{eq:9}
  [U^m\partial_x \Udag^m, U^m\partial_y \Udag^m] \left(\rme^{\rmi m\phi}\psi\right) = 0.
\end{equation}
The partial derivatives transform according to
\begin{equation}
  \label{eq:10}
  U^m\partial_x \Udag^m = \partial_x - \rmi m \partial_x \phi, \quad
  U^m\partial_y \Udag^m = \partial_y - \rmi m \partial_y \phi.
\end{equation}
Substituting (\ref{eq:10}) into (\ref{eq:9}), and using the
differential version of Stokes' theorem,
\begin{equation}
  \label{eq:11}
  [\partial_x,\partial_y]\phi = (\nablavec\wedge\nablavec)_z \phi = 2\pi \delta^2(\rvec),
\end{equation}
we obtain
\begin{equation}
  \label{eq:12}
 [\partial_x,\partial_y] \left(\rme^{\rmi m\phi} \psi\right) = 2\pi   \delta^2(\rvec)
 \rmi m\rme^{\rmi m\phi} \psi
= 2\pi \delta^2(\rvec) \partial_\phi \left(\rme^{\rmi m \phi} \psi \right),
\end{equation}
where the second equality follows because $\partial_\phi\psi$ 
vanishes at the
origin for $\psi(\rvec)$ smooth.  This implies, formally, the operator relation
\begin{equation}
  \label{eq:18}
  [\partial_x,\partial_y] = 2 \pi \delta^2(\rvec) \partial_\phi,
\end{equation}
or, equivalently,
\begin{equation}
  \label{eq:4}
  [P_x, P_y] = -2\pi \rmi \hbar \delta^2(\rvec) L,
\end{equation}
where $L = (\hbar/\rmi) \partial_\phi$ is the canonical angular
momentum.

Using the modified commutation relation (\ref{eq:4}) to evaluate the
magnetic field operator (\ref{eq:2}),
we get, instead of the classical relation $B = \alpha \Phi_0
\delta^2(\rvec)$,  the result
\begin{equation}
  \label{eq:15}
  B = \frac{\Phi_0}{2\pi \rmi \hbar^2}[P_x,P_y] +
  (\nabla\wedge \Avec)_z 
 = -\frac{\Phi_0}{\hbar} \delta^2(\rvec)\Lambda,
\end{equation}
where
\begin{equation}
  \label{eq:7}
  \Lambda\zhat = \rvec \wedge M\Vvec =
  (L - \alpha\hbar)\zhat
\end{equation}
is the kinetic angular momentum.  It is evident that $\Lambda$, and
hence $B$, satisfy the transformation law (\ref{eq:6}).

Throughout this and the following sections, it will be convenient to
represent vectors in the $xy$-plane as complex scalars. For
example, $\Wvec = W_x\xhat + W_y\yhat$ will be represented by $\Wcal = W_x
+ \rmi W_y$.  If $\Wvec$ is a vector of hermitian operators (as opposed to
real scalars), then $\Wcal$ is the nonhermitian scalar operator whose
hermitian and antihermitian parts are $W_x$ and $\rmi W_y$
respectively.  In this way, the kinetic momentum $M\Vvec$ is
represented by the scalar operator
\begin{equation}
  \label{eq:13}
  M\Vcal = \frac{\hbar}{\rmi} \rme^{\rmi \phi} \left(\partial_r + \frac{\rmi
  \partial_\phi + \alpha}{r}\right).
\end{equation}
It is useful to note that, in general, $\zhat \wedge \Wvec$ is
represented by $\rmi \Wcal$.

We proceed to compute the expectation value of the Lorentz force.
Substituting (\ref{eq:15}) for the magnetic field
and  (\ref{eq:13})  for the kinetic momentum into the expression
(\ref{eq:1}), we get 
\begin{eqnarray}
  \label{eq:14} 
  \fl \me{\psi}{\Fcal}{\psi} = -\rmi \frac{e}{c} \me{\psi}{B\Vcal}{\psi}\nonumber\\
  \fl\qquad           = -\frac{2\rmi\hbar^2}{M}\int_0^{2\pi}\rmd\phi\, \int_0^\infty  
                  \psi^*(r,\phi)\left[\frac{\delta(r)}{r}(\partial_\phi - \rmi\alpha)
                    \rme^{\rmi \phi}\left(\partial_r
                      +\frac{\rmi\partial_\phi +
                        \alpha}{r}\right)\right]\psi(r,\phi)r\,\rmd r ,
\end{eqnarray}
where we have used $\delta^2(\rvec) = \delta(r)/(\pi r)$.  
With $\psi$ resolved into its angular momentum components,
\begin{equation}
\label{eq:decompose}
\psi(r,\phi) = \sum_{m=-\infty}^{\infty} \psi_m(r) \rme^{\rmi m \phi},
\end{equation}
the integrals in (\ref{eq:14}) are trivially evaluated (note that $\int_0^\infty \delta(r) \,\rmd r
= \case12$), with the result
\begin{equation}
  \label{eq:f2}
\fl  \me{\psi}{\Fcal}{\psi} = \frac{2\pi\hbar^2}{M}
  \sum_{m=-\infty}^{\infty}
  \left[\psi_{m+1}^*(r)(m+1-\alpha)\left(\psi_m'(r) - 
  \frac{(m-\alpha)}{r}\psi_m(r)\right)\right]_{r = 0}.
\end{equation}
Like any vector operator, $\Fcal$ couples only consecutive angular
momentum components, $m$ and $m+1$, and, as one would expect, depends only on the behaviour of the
wavefunction at the flux line.  For well-behaved wavefunctions, the leading-order
behaviour of $\psi_m(r)$ as $r \rightarrow 0$ is given by
\begin{equation}
  \label{eq:r=0}
  \psi_m(r) \sim C_m r^{|m - \alpha|}.
\end{equation}
Sufficient conditions for
(\ref{eq:r=0}) are discussed in the Appendix.  Here, we note that
(\ref{eq:r=0}) ensures that the energy density,
$\psi^*(\rvec)(H\psi)(\rvec)$, is finite at $r = 0$.
%insures that energy density, $\psi^*(\rvec)(H\psi)(\rvec)$,
%is nonsingular at $r = 0$, since
%the Hamiltonian acts on the angular momentum components
%according to 
%\begin{equation}
%  \label{eq:Hpsi}
%  H (\psi_m \rme^{\rmi m\phi}) = -\frac{\hbar^2}{2M} \left(\psi_m'' +
%  \frac{1}{r} {\psi_m'} + \frac{(m-\alpha)^2}{r^2}\psi_m\right) \rme^{\rmi m\phi},
%\end{equation}
%(The wavepackets considered in the Appendix necessarily satisfy (\ref{eq:r=0})).
Substituting (\ref{eq:r=0}) into (\ref{eq:f2}), we get
\begin{eqnarray}
  \label{eq:f2two}
\fl  \me{\psi}{\Fcal}{\psi} = \frac{2\pi\hbar^2}{M}
  \sum_{m=-\infty}^{\infty}  (m+1-\alpha)(|m-\alpha| -
  (m-\alpha)) \nonumber\\
\lo \times \left.C^*_{m+1} C_m r^{|m+1-\alpha| + |m-\alpha| - 1}\right|_{r = 0}.
\end{eqnarray}

The terms in the sum (\ref{eq:f2two}) vanish unless  $m = a$, 
where $a = \alpha - \fracalph$ denotes the integer part of the flux parameter. Thus, only
the `pinioned' components of the wavefunction, $\psi_a$ and $\psi_{a+1}$ --
those whose angular momentum quantum numbers are nearest the flux parameter
$\alpha$ -- contribute to the force expectation value (\ref{eq:f2}).
From~\eqref{eq:f2two},
\begin{equation}
  \label{eq:fexpect}
   \me{\psi}{\Fcal}{\psi} = \frac{4\pi\hbar^2}{M}\fracalph(1-\fracalph)C^*_{a+1}C_a,
\end{equation}
or, equivalently,
\begin{equation}
  \label{eq:forceop}
  \Fcal = \frac{4\pi\hbar^2}{M}\fracalph(1-\fracalph)\ket{\xi_{a+1}}\bra{\xi_a},
\end{equation}
where the state $\ket{\xi_m}$ corresponds to the singular wavefunction
\begin{equation}
  \label{eq:xi}
  \xi_m(r,\phi) = \frac{\delta(r)}{\pi r^{|m - \alpha| + 1}} \rme^{\rmi m \phi},
\end{equation}
so that $\braket{\xi_m}{\psi} = C_m$.
It is readily verified that the force operator (\ref{eq:forceop})
transforms according to~\eqref{eq:6} under the gauge transformation~\eqref{eq:5}.
A rigorous derivation of the expectation value (\ref{eq:fexpect}) for
suitably chosen wavefunctions is given in the Appendix.

In the preceding derivation, the force operator due to a flux line,
like the force for a nonsingular potential, is derived from kinematics,
specifically from the commutation relations. The derivation does not
require the solution of the Schr\"odinger equation.  Thus, it is
straightforward to generalize to the case of more than one flux line
(for which solutions of the Schr\"odinger equation are, in 
general, not available);
the force operator is just a sum of contributions (\ref{eq:forceop})
centred around each flux line.

However, to calculate the force on stationary states, or the
time dependence of the force on nonstationary states, it is necessary
to solve the Schr\"odinger equation.  As is well known, eigenstates of
the Aharonov-Bohm Hamiltonian with energy $\hbar^2k^2/2M$ and angular
momentum $m\hbar$ are given by
\begin{equation}
  \label{eq:eigen}
  \chi_{k,m}(\rvec) = J_{|m -\alpha|}(kr) \rme^{\rmi m \phi},
\end{equation}
where $J_\nu(z)$ is a Bessel function.  From the small-$z$ behaviour, $J_\nu(z)
\sim (z/2)^\nu/\Gamma(\nu + 1)$, and the reflection
formula, $\Gamma(\nu)\Gamma(1-\nu) = \pi/\sin\pi\nu$, we obtain from
(\ref{eq:forceop}) the matrix elements
\begin{equation}
  \label{eq:matrix}
  \me{\chi_{p,n}}{\Fcal}{\chi_{k,m}} =
  \frac{2\hbar^2}{M} k^\fracalph
  p^{1-\fracalph}\sin\pi\fracalph\,\delta_{m,a} \delta_{n,a+1}.
\end{equation}
For $k = p$, ie for stationary states with the same
energy, we obtain the on-shell matrix elements
\begin{equation}
  \label{eq:onshellmatrix}
  \me{\chi_{k,n}}{\Fcal}{\chi_{k,m}} =
  \frac{2\hbar^2}{M} k
\sin\pi\fracalph\,\delta_{m,a} \delta_{n,a+1},
\end{equation}
in agreement with results of
Nielson and Hedeg\aa rd (1995) and Shelankov (2000).

A general stationary state $\ket{\Psi}$ is a superposition of eigenstates with
$k$ fixed, and may be taken to be of the form
\begin{equation}
  \label{eq:super}
  \ket{\Psi} = \sum_{m=-\infty}^{\infty} (-\rmi)^{|m-\alpha|} b_m \ket{\chi_{k,m}}.
\end{equation}
For $b_m = 1$, $\Psi(\rvec)$ corresponds to a scattered plane wave incident from the
right (Aharonov \& Bohm 1959).  
From~\eqref{eq:onshellmatrix}, we get the expectation value
\begin{equation}
  \label{eq:stationary}
  \me{\Psi}{\Fcal}{\Psi} = \frac{2\rm i\hbar^2}{M} \rme^{-\rmi \pi\fracalph}k\sin\pi\fracalph\, 
   b^*_{a+1} b_{a}.
\end{equation}
%or, equivalently,
%\begin{equation}
%  \label{eq:cartesian}
%  \me{\Psi}{\Fvec}{\Psi} = \frac{2\rm i\hbar^2\sin\pi\fracalph}{M} 
%  \left|b_{a+1}b_a\right| k \left( \sin(\pi\fracalph + \delta)\xhat + 
%    \cos(\pi\fracalph + \delta)\yhat\right)
%\end{equation}
%where $\delta = \arg b_{a+1} - \arg b_a$.  This generalizes results of
%Nielson and Hedegard (1995), Shelankov (2000), etc, whose results
%yield expectation values and matrix elements of the force for
%stationary states with the same energy.

Shelankov (1998) has obtained an approximate formula for the
transverse component of the force acting on a stationary beam of
finite angular width.  His analysis is carried out in a singular
gauge, in which the vector potential vanishes everywhere except along
the $y$-axis.  A stationary beam incident from the right
(in fact, Shelankov takes the beam incident from the left, but we
revert to the convention of Aharonov and Bohm) is taken to be of the form $\rme^{-\rmi k
  x}\psi(x,y)$. Treating $\psi_{xx}$ as small compared to $k\psi_x$
amounts to a paraxial approximation, in which the wave evolves freely
in $x$ (with $x$ playing the role of time) for $x \ne 0$, and is
scattered by the vector potential at $x = 0$.  The change $\Delta p_y$
in the transverse kinematic momentum, $(\hbar/\rmi) \int_{-\infty}^\infty
 \psi^*(x,y)\psi_y(x,y)\,\rmd y$, is then calculated to be
\begin{equation}
  \label{eq:shelimpulse}
  \Delta p_y = \hbar \sin 2\pi\alpha\frac{\left|\psiin(0)\right|^2}{\int_{-\infty}^{\infty}
  \left|\psiin(y)\right|^2 \,\rmd y} 
\end{equation}
where
\begin{equation}
  \label{eq:psiin}
  \psiin(y) = \frac{1}{\sqrt{2\pi}k}\int_{-\infty}^{\infty}a(k_y) \rme^{\rmi
  k_y y} \,  \rmd k_y 
\end{equation}
is the incident wave at $x = 0^+$, expressed here in terms of its transverse
Fourier amplitudes $a(k_y)$.  Multiplying $\Delta p_y$ by the incident
flux, which is given paraxially by $(\hbar
k/M)\int_{-\infty}^{\infty}  \left|\psiin(y)\right|^2 \,\rmd y$, gives {\it Shelankov's
formula} for
the transverse force,
\begin{equation}
  \label{eq:shelforce}
    F_y^{(S)} = \frac{\hbar^2}{M}k\sin 2\pi\fracalph \left|\psiin(0)\right|^2.
\end{equation}

We now show that the $y$-component of the exact force expectation
value, ie the imaginary part of (\ref{eq:stationary}), coincides with
Shelankov's formula (\ref{eq:shelforce}) in the paraxial regime.
(Shelankov (2000) gives the same argument.) As discussed by Berry
(1999), the state (\ref{eq:super}) can alternatively be viewed as a
superposition of scattered waves incident from the directions
$(\cos\theta,-\sin\theta)$, with amplitudes $A(\theta)$ related to the
coefficients $b_m$ according to
\begin{equation}
  \label{eq:b_m}
  b_m = \frac{1}{\sqrt{2\pi}} \int_{-\pi}^{\pi}  A(\theta)
  \rme^{\rmi (m - \alpha)\theta}\, \rmd \theta.
\end{equation}
The paraxial approximation is valid for $A(\theta)$ strongly
peaked around $\theta = 0$, with angular width $w << 1$.  In this case,
Berry (1999) has shown that $A(\theta)\sim a(k\theta)$. 
From (\ref{eq:psiin}) and (\ref{eq:b_m}),
it then follows that $b_m \sim \psiin((m-\alpha)/k)$ for $|m - \alpha| <<
1/w$, so that $b^*_{a+1} b_a \sim |\psiin(0)|^2$.

%In a recent preprint, Shelankov (2000) gives a nice derivation of the force matrix
%elements (\ref{eq:matrix}) and force expectation value
%(\ref{eq:stationary}) for stationary states, and shows, using Berry's
%asymptotics as above, that the transverse component agrees with his
%paraxial formula (\ref{eq:shelforce}) in the appropriate regime.  His
%derivation is based on the stress tensor...

\section{Impulse operator}\label{sec:impulse-operator}
For nonstationary wavepackets $\psi(\rvec)$, whose wavefunctions 
are not eigenfunctions of the Hamiltonian, the
expectation value of the force does not itself have much physical
significance.  It depends on the behaviour of the wavefunction near
the flux line, regardless of where the wavepacket is localized, and
can oscillate rapidly as the wavepacket evolves.  Of greater physical
interest is the impulse imparted to the particle over the course of
its evolution, either in the past or future.  Let 
\begin{equation}\label{eq:time-evolved}
\Fcal(t) =  \rme^{\rmi H t/\hbar} \Fcal \rme^{-\rmi H
t/\hbar},\qquad
M\Vcal(t) =  \rme^{\rmi H t/\hbar} M\Vcal \rme^{-\rmi H
t/\hbar}
\end{equation}
denote the time-evolved force and kinetic momentum operators.  Then
$M\dot\Vcal(t) = \Fcal(t)$.
The forward (+) and backward
(-) impulse operators are defined by
\begin{equation}
  \label{eq:impulse}
  \Icalpm = M\Vcal(\pm\infty) - M\Vcal(0) = 
  \int_0^{\pm\infty}\Fcal(t)\,\rmd t.
\end{equation}

Let us compute the kernal of the impulse operator in the position representation,
$\Ical_\pm(\svec,\rvec) = \me{\svec}{\Ical_\pm}{\rvec}$.
From the
completeness relation,
\begin{equation}
  \label{eq:res}
 \frac{1}{2\pi} \sum_{m = -\infty}^{\infty}\int_0^\infty
  \ket{\chi_{k,m}}\bra{\chi_{k,m}}k\,\rmd k = 1,
\end{equation}
we obtain
\begin{eqnarray}
  \label{eq:res2a}
\fl \Icalpm(\svec,\rvec) =
\frac{1}{4\pi^2}\sum_{m=-\infty}^{\infty}\sum_{n=-\infty}^{\infty}\int_0^{\pm\infty}\rmd
t \nonumber\\
\fl\quad \times \int_0^\infty\!\!\int_0^\infty 
\braket{\svec}{\chi_{p,n}}\me{\chi_{p,n}}{\Fcal }{\chi_{k,m}}
\braket{\chi_{k,m}}{\rvec}\exp(\rmi\hbar(p^2-k^2)t/2M)
\,k\rmd k \, p\rmd p.
\end{eqnarray}
From 
the expression (\ref{eq:matrix}) for the matrix elements $\me{\chi_{p,n}}{\Fcal }{\chi_{k,m}}$,
the only contribution to the double sum in~\eqref{eq:res2a} is from the term $n = a+1$, $m
= a$.  Substituting the eigenfunctions~\eqref{eq:eigen}, and 
letting $(s,\theta)$ denote the polar coordinates of $\svec$, we get
\begin{eqnarray}\label{eq:res2}
 \fl \Icalpm(\svec,\rvec) = \pm \frac{\hbar^2 \sin\pi\fracalph}{2\pi^2 M}\exp(\rmi(a+1)\theta -
   \rmi a \phi)\int_0^\infty \rmd t\nonumber\\
\fl\qquad \times  
\int_0^\infty\exp\left(\pm \rmi \frac{\hbar p^2}{2M}t\right) J_{1-\fracalph}(ps)
  p^{2-\fracalph}\, \rmd p
\int_0^\infty \exp\left(\mp \rmi \frac{\hbar k^2}{2M}t\right) J_\fracalph(kr)
  k^{1+\fracalph}\,  \rmd k.
\end{eqnarray}
The $k$- and
$p$-integrals are of the form (Abramowitz \& Stegun 1970)
\begin{equation}
  \label{eq:integral}
  \int_0^\infty \rme^{-c^2 u^2} J_\nu(bu)  u^{\nu + 1}\,\rmd u =
  \frac{b^{\nu}}{(2c^2)^{\nu+1}}\rme^{-b^2/4c^2}.
\end{equation}
Substituting this result into~\eqref{eq:res2}, we get
\begin{eqnarray}
  \label{eq:i3}
\fl \Icalpm(\svec,\rvec) = \frac{\rmi M^2 }{2\pi^2\hbar}\sin\pi\fracalph\,
  \exp(\rmi(a+1)\theta - \rmi a\phi \mp\rmi\pi\fracalph)
   s^{1-\fracalph} r^{\fracalph}\nonumber\\
\times \int_0^\infty 
  \exp\left(\mp \rmi \frac{M (r^2-s^2)}{2\hbar t}\right) t^{-3}\,  \rmd t.
\end{eqnarray}
With the substitution $t = 1/w$, the remaining integral in
(\ref{eq:i3}) is of the
elementary form
\begin{equation}
  \label{eq:t-int}
  \lim_{\epsilon\rightarrow0^+}
\int_0^\infty \exp\left(-\frac{\epsilon \pm\rmi (r^2-s^2)w}{\sigma^2}\right) w \, \rmd w=
\frac{\sigma^4}{(0^+ \pm \rmi (r^2-s^2))^2},
\end{equation}
with $\sigma^2 = 2\hbar/M$.  Substituting (\ref{eq:t-int}) into
(\ref{eq:i3}), we obtain
\begin{equation}
  \label{eq:ifinal}
\fl  \Icalpm(\svec,\rvec) =
  \frac{2\rmi\hbar}{\pi^2}\sin\pi\fracalph\,
\exp(\rmi(a+1)\theta - \rmi a\phi \mp\rmi\pi\fracalph)
\frac{r^\fracalph
    s^{1-\fracalph}}{(0^+ \pm \rmi (r^2-s^2))^2}.
\end{equation}
One can verify that this expression
transforms correctly, ie according to (\ref{eq:6}), under the gauge
transformation
(\ref{eq:5}).

The denominator in (\ref{eq:ifinal}) can be alternatively expressed as
\begin{equation}
  \label{eq:denom}
  \frac{1}{(0^+ \pm \rmi (r^2-s^2))^2}
 = P'(1/(r^2-s^2)) \pm \rmi\pi\delta'(r^2-s^2).
\end{equation}
Here $P'(1/x)$, the derivative of the principal part, acts on
functions $f(x)$ according to
\begin{equation}
  \label{eq:principal}
  \int_{-\infty}^\infty f(x)P'(1/x)\,\rmd x  = -
  \int_{-\infty}^{\infty}
  f_{\rm odd}'(x)/x\,\rmd x,
\end{equation}
where $f'_{\rm odd}(x) = \case12(f'(x)-f'(-x))$ denotes the odd part of $f'(x)$.  For
subsequent calculations, however, the integral representation
(\ref{eq:t-int}) will be more convenient.

\section{Expectation values of impulse for wavepackets}\label{sec:Iexpectation}

We parameterize wavepackets by their position $\Rvec$, width $\sigma$ and
kinetic momentum $\hbar\kvec$.  A convenient form is 
\begin{equation}
  \label{eq:psi}
  \psi(\rvec) = \frac{1}{\sigma}
  f\left(\frac{\rvec-\Rvec}{\sigma}\right) \exp(-\rmi
  \kvec\cdot\rvec + \rmi \alpha\phi).
\end{equation}
Here $f(\uvec)$ is a smooth normalized function localized at the
origin with unit width and vanishing average (dimensionless) momentum, ie
\begin{equation}
  \label{eq:fprops}
\fl  \int\!\!\int f^*f \, \rmd^2 u  = 1,
  \int\!\!\int   f^* f  \uvec\, \rmd^2 u = \zerovec, \quad 
  \int\!\!\int f^*  f  u^2 \,\rmd^2 u = 1, \quad
  \int\!\!\int f^*\nablavec f\,  \rmd^2 u = \zerovec.
\end{equation}
We assume that $\sigma<< R$,
so that the wavepacket $\psi(\rvec)$ is localized far from the flux
line.  The phase factor $\exp(\rmi\alpha\phi)$ in (\ref{eq:psi})
insures that $\hbar k$  is the kinetic, rather than the canonical,
momentum of the wavepacket; its branch is chosen so that the phase factor is
continuous over the region where $\psi(\rvec)$ is appreciable.

From  
(\ref{eq:ifinal}) and (\ref{eq:psi}), the
expectation value of the impulse
is given by
\begin{eqnarray}
  \label{eq:iexpect2a}
\fl  \Iexpect(\Rvec,\kvec,\sigma,\alpha) = \me{\psi}{\Icalpm}{\psi} = 
\frac{2\rmi\hbar}{\pi^2\sigma^2}\rme^{\mp\rmi\pi\fracalph}\sin\pi\fracalph
\nonumber\\
\fl\quad  \times \int\!\!\int\!\!\int\!\!\int f^*\left(\frac{\svec-\Rvec}
{\sigma}\right) f\left(\frac{\rvec-\Rvec}{\sigma}\right)
\frac{\rme^{\rmi \kvec\cdot(\rvec-\svec)}}{(0^+ \pm \rmi (r^2-s^2))^2}
\rme^{
\rmi (1-\fracalph)\theta + \rmi
    \fracalph \phi}r^\fracalph s^{1-\fracalph}\,\rmd^2r\,\rmd^2s.
\end{eqnarray}
Since $f$ has unit width, the integrand in~\eqref{eq:iexpect2a}
is appreciable only for 
$|\svec - \Rvec|\sim \sigma$ and $|\rvec - \Rvec|\sim \sigma$.  In
this region, we can, to leading order in 
$\sigma/R$, replace the phase factors $\exp(\rmi (1-\fracalph)\theta)$ and $\exp( \rmi
    \fracalph \phi)$   by $\exp(\rmi (1-\fracalph)\Phi)$ and  $\exp( \rmi
    \fracalph \Phi)$, respectively, where 
$\Phi$ is the polar
    angle of  $\Rvec$.  Likewise, we can replace the factor
    $r^\fracalph s^{1-\fracalph}$ by $R$.
With the change of variables
$\uvec = (\rvec - \Rvec)/\sigma$ and $\vvec = (\svec - \Rvec)/\sigma$
and the integral representation
(\ref{eq:t-int}), \eqref{eq:iexpect2a} becomes 
\begin{eqnarray}
  \label{eq:K2}
\fl \Iexpect = \frac{2\rmi\hbar}{\pi^2}\frac{R}{\sigma^2}\rme^{\rmi \Phi \mp\rmi\pi\fracalph}
\sin\pi\fracalph\nonumber\\
\lo \times \int_0^\infty
 \rmd w\, w \left|
\int\!\!\int f(\uvec)
\exp\left(\rmi\sigma\kvec\cdot\uvec \mp 2\rmi w\frac{\Rvec\cdot\uvec}{\sigma} \mp\rmi
 w u^2 \right)\,\rmd^2 u\right|^2.
\end{eqnarray}
Thus, the
direction of the impulse, $\arg \Iexpect$, is given by
\begin{equation}
  \label{eq:direction}
  \arg \Iexpect=  \Phi +
(\case12 \mp \fracalph)\pi.
\end{equation}
For $\fracalph = \case12$, the forward
impulse is directed away from the flux line, and the backwards impulse
towards the flux line.   

There are two parameter regimes where the
expression~\eqref{eq:K2} has a simple asymptotic form, namely $k\sigma << 1$,
which corresponds to slow wavepackets, and $k\sigma >> R/\sigma$,
which corresponds to fast wavepackets.  These cases are discussed
separately below.

\subsection{Slow wavepackets.}\label{sec:slow}
The condition $k\sigma << 1$ implies that the wavepacket spreads
(with velocity $\sim \hbar/M\sigma$) more quickly than it moves (with
velocity $\hbar k/M$). Since $f$ has unit width, the integrand
in (\ref{eq:K2}) is appreciable only for $u$ of order one.  In this
case, for $k\sigma << 1$, the phase factor $\exp (\rmi
\sigma\kvec\cdot \uvec)$ is nearly equal to one.
On the other hand, the phase factor $\exp ( \mp 2 \rmi
w\Rvec\cdot \uvec/\sigma)$ oscillates rapidly in this region, and 
hence renders the integral small, unless $w$ is small, of order
$\sigma/R$.  For $w$ of order $\sigma/R$, the phase factor $\exp (\rmi
u^2 w)$ is nearly equal to one for $u$ of order one.
To
leading order in $\sigma/R$ and $k\sigma$, (\ref{eq:K2}) becomes
\begin{eqnarray}
  \label{eq:slowa}
  \Iexpect &=&   \frac{2\rmi\hbar}{\pi^2}\frac{R}{\sigma^2}\rme^{\rmi \Phi \mp\rmi\pi\fracalph}
\sin\pi\fracalph\,
 \int_0^\infty
\rmd w\,w
\left|\int \!\!\int f(\uvec) \rme^{\mp 2\rmi w\Rvec\cdot \uvec /\sigma}
 \, \rmd^2 u\right|
\nonumber\\ 
&=& 8\rmi\hbar\frac{R}{\sigma^2}\rme^{\rmi \Phi \mp\rmi\pi\fracalph}
\sin\pi\fracalph\,
\int_0^\infty
\left|\ftilde\left(\pm\frac{2\Rvec}{\sigma}w\right)\right|^2 w \, \rmd w,
\end{eqnarray}
where 
\begin{equation}\label{eq:nft}
\ftilde(\xivec) = 
\frac{1}{2\pi}\int\!\!\int  f(\uvec) \rme^{-\rmi \xivec\cdot\uvec}\, \rmd^2 u
\end{equation}
denotes the normalized Fourier transform of $f(\uvec)$.
Letting
\begin{equation}
  \label{eq:rho}
  \rhotilde(\ehat) = \int_0^\infty   \left|\ftilde(\xi\ehat)\right|^2\xi\, \rmd \xi
\end{equation}
denote the probability distribution for the direction, $\ehat$, of the
dimensionless momentum, $\xivec = \xi\ehat$, we can write
\begin{equation}
\label{eq:slow}
\Iexpect = \frac{2\rmi\hbar}{R}
\rme^{\rmi \Phi \mp\rmi\pi\fracalph}\sin\pi\fracalph\,
\rhotilde(\pm\Rhat).
\end{equation}
Note that 
if $f(\uvec)$ is
circularly symmetric, then $\rhotilde(\ehat)$ is equal to $1/2\pi$.

%to leading order in
%, the expression
%\begin{equation}
%  \label{eq:slow2}
%  \me{\psi}{\Icalpm}{\psi} = 2\rmi \frac{\hbar}{R} \exp(\rmi (\Phi \mp
%  \fracalph\pi))\sin(\pi\fracalph)\rhotilde(\pm\Rhat)
%\end{equation}
%%or its cartesian equivalent,
%for the impulse imparted to a slow wavepacket.  

To leading order in $\sigma/R$ and $k\sigma$, the
impulse~\eqref{eq:slow} is independent of the width and momentum of the wavepacket,
and is of order $\hbar/R$ (ie, inversely proportional to the
distance from the flux line).  This is much smaller than the
dispersion of the momentum, which is of order $\hbar/\sigma$.
Therefore, to detect the impulse on slow wavepackets experimentally,
one would have to perform a large number of measurements (on the order
of $(R/\sigma)^2$) of the asymptotic momentum on an ensemble of
identically prepared systems.
 
By treating the motion of the centre of a slow wavepacket as a classical
trajectory, we can derive an expression for the scattering
cross-section $\sigma(\theta)$ using the classical formula,
\begin{equation}\label{eq:classicalxsec}
\sigma(\theta) = \left | \frac{\rmd b}{\rmd \theta}(\theta)\right|.
\end{equation}
Here $b$ is the impact parameter, and $\theta$ is the angular
direction of the outgoing trajectory.
Consider a slow wavepacket (\ref{eq:psi})
centred on the $y$-axis at $Y\yhat$ at $t=0$ (thus, $R = |Y|$ and
$\Phi = \sgn (Y) \pi/2$),
moving in the $-\xhat$ direction with kinetic momentum $\hbar k$. For simplicity, we take $f(\uvec)$
to be circularly symmetric, so that $\rhotilde = 1/2\pi$.  
Let $\beta_-$ denote the angle
between the velocities at $t= -\infty$ and $t = 0$, and $\beta_+$ the
angle between the velocities at $t= 0$ and $t = \infty$ (see Fig~1).
From (\ref{eq:slow}), these are given by
\begin{equation}
  \label{eq:betapm}
  \cot\beta_\pm = \pm \frac{\re \me{\psi}{\Ical_\pm}{\psi} - \hbar
    k}
{\im \me{\psi}{\Ical_\pm}{\psi}}
                =  \frac{\sin \pi \fracalph \cos \pi\fracalph -
                  \pi kY
                  }{\sin^2 \pi\fracalph}
\end{equation} 
(thus $\beta_- = \beta_+$).
\begin{figure}\label{Fig1}
\begin{center}
\input{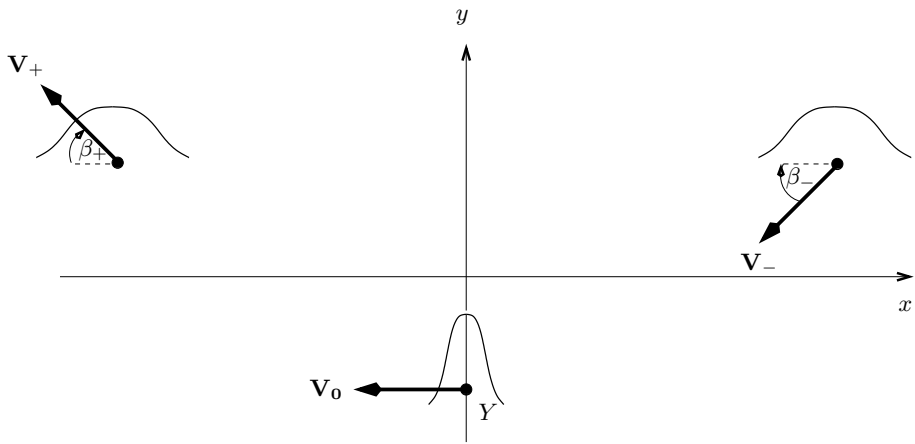}
\caption{At $t = 0$ the wavepacket is centred at $Y\yhat$ and moves in the $-\xhat$ direction.
$\beta_\pm$ are the angles of the incoming and outgoing velocities with respect to $-\xhat$.}
\end{center}
\end{figure}
Because the impulse is circularly symmetric, the angles $\beta_\pm$ are
unchanged if we rotate the entire system so that the incoming velocity, at $t = -\infty$, is 
in the $-\xhat$ direction.  In this case, the direction of the
outgoing beam is given by
\begin{equation}
  \label{eq:theta=}
  \theta = \pi + \beta_+ + \beta_- = \pi + 2\beta_+.
\end{equation}

To determine the impact parameter $b$, we appeal to classical
angular momentum conservation, 
$MV_{-} b =
\sgn b\, MV_{\rm 0} R_0$,
where $V_{-}$ is the speed at $t = -\infty$, and $R_0$ and
$V_0$ are the distance and speed at the point of closest approach to
the flux line.
For the Aharonov-Bohm
Hamiltonian (and, indeed, for any purely magnetic Hamiltonian), the
speed $V = \sqrt{\Vvec\cdot \Vvec}$ is a conserved quantity.  Thus
$b = \sgn b\, R_0$. We take the point of closest approach 
to occur at $t = 0$ (when the velocity of the wavepacket is orthogonal to its
position), so that $b = Y$.  
Then, from (\ref{eq:classicalxsec}), (\ref{eq:betapm}) and
(\ref{eq:theta=}),  
\begin{eqnarray}
  \label{eq:xsection}
   \sigma(\theta) &=& \left|\frac{\rmd b}{\rmd  \theta}\right| 
= \left|\frac{\rmd \theta}{\rmd b}\right|^{-1}
\left|
  2\frac{\rmd \beta_+}{\rmd Y}  
\right|^{-1} = 
\left|
2  \sin^2 \beta_+\frac{\rmd (\cot \beta_+)}{\rmd Y} 
\right|^{-1}\nonumber\\
&=& \frac{\sin^2\pi\fracalph}{2\pi k \cos^2 \theta/2}.
\end{eqnarray}

Surprisingly, the expression (\ref{eq:xsection}) agrees with 
the exact result found by Aharonov \& Bohm (1959).  Of
course, the preceding should not be regarded as a
legitimate derivation of the scattering cross-section.  Apart from certain
{\it ad hoc} elements (eg, circularly
symmetric wavepacket and the determination of the impact parameter), a proper
derivation of the cross-section from time-dependent solutions of the
Schr\"odinger equation requires wavepackets (unlike the slow ones used
here) whose momentum is sharp.  Still, it is interesting to ask
whether or not this agreement is purely fortuitous.

\subsection{Fast wave packets}\label{sec:fast-wave-packets}
A wavepacket initially at a distance $R$ from the flux line with
kinetic momentum $\hbar k$ reaches its point of closest approach to
the flux line after a time $\tau$ of order $R/(\hbar k/M)$.  It
spreads at a speed
$W$ of order 
$\hbar/M\sigma$.
Thus, at closest approach it
will have spread a distance of order $W\tau\sim  R/k\sigma$.  For
this to be much less than the width $\sigma$, we require $R << k \sigma^2$,
which is just the condition for fast wavepackets.

Let
\begin{equation}
  \label{eq:uvecint}
  L_\pm = \left|\int\!\!\int f(\uvec)
\exp\left(\rmi\sigma\kvec\cdot\uvec \mp 2\rmi w\Rvec\cdot\uvec /\sigma \mp\rmi
 w u^2\right)\,\rmd^2 u\right|^2
\end{equation}
denote the $\uvec$-integral which appears in the impulse expectation value 
(\ref{eq:K2}).  Because
$f$ is of unit width, the integrand is appreciable only for $u$ of
order $1$.
For $k\sigma >> R/\sigma$, the phase factor $\exp (\rmi \sigma
\kvec\cdot\uvec)$ 
in (\ref{eq:K2}) is rapidly oscillating,
and hence
leads to a vanishingly small integral unless it is balanced by
the phase factor $\exp (\mp \rmi 2iw\Rvec\cdot\uvec/\sigma)$.  For such
a balancing to take place, $w$ must be large, of order $k\sigma^2/R$.
Therefore, the quadratic phase factor $\exp (\mp \rmi w u^2)$ is rapidly
oscillating, so
(\ref{eq:uvecint}) can be evaluated using the stationary phase approximation.  To
leading order in $1/w \sim R/(k\sigma^2)$, we obtain
\begin{equation}
  \label{eq:Iu}
  L_\pm = \frac{\pi^2}{ w^2} \left|f\left(\frac{\mp \sigma^2 \kvec/2w
 - \Rvec}{\sigma}\right)\right|^2.
\end{equation}
Substituting into~\eqref{eq:K2}, we get
\begin{eqnarray}
  \label{eq:Kfast}
\Iexpect  &=& 2\rmi\hbar\frac{R}{\sigma^2}\rme^{\rmi \Phi \mp\rmi\pi\fracalph}
\sin\pi\fracalph\,
 \int_0^\infty
 \left| f\left(\frac{\mp \sigma^2 \kvec/2w
 - \Rvec}{\sigma}\right)\right|^2\, \frac{\rmd w}{w}\nonumber\\
&=&  2\rmi\hbar R \rme^{\rmi \Phi \mp\rmi\pi\fracalph}
\sin\pi\fracalph\,
 \int_0^\infty \left|\psi(\mp r\khat)\right|^2\, \frac{\rmd r}{r},
\end{eqnarray}
where we have used (\ref{eq:psi}) to express the integral in terms of
the wavefunction $\psi(\rvec)$.  Note that $\psi(\rvec)$
behaves for small $r$ as $r^\fracalph$ or
$r^{1-\fracalph}$ (cf (\ref{eq:r=0})), so that the integral in
(\ref{eq:Kfast}) is
convergent.

In what follows, let us assume for concreteness that $\kvec$ is
directed along -$\xhat$, so that the wavepacket is moving to the left.
We write $\Rvec = X\xhat + Y\yhat$. Unless the wavepacket is centred
near the $x$-axis (specifically, 
unless $|Y|\sim \sigma$), $\psi(\mp r\khat)$  will be negligible over the range of integration
in~\eqref{eq:Kfast}.
Thus, to leading order in $\sigma/R$, we may take $R\rme^{\rmi\Phi} = X$.
Substituting this result into \eqref{eq:Kfast}, we obtain the expression
\begin{equation}
  \label{eq:Ifastaa}
\Iexpect = 2\rmi \hbar
\rme^{\mp\rmi\pi\fracalph}X\sin\pi\fracalph\,
 \int_0^{\pm\infty} \left|\psi(x,0)\right|^2\, \frac{\rmd x}{x}.
\end{equation}
for the impulse.

Since the wavepacket is centred near $X\xhat$, the $x$-integral in~\eqref{eq:Ifastaa} is negligible unless
$X > 0$ in the forward $(+)$ case
(so that the wavepacket is moving towards 
the flux line), or unless $X < 0$  in the backward (-) case (so that the wavepacket is
moving away from the flux line).
Assuming that $\pm X > 0$, 
the main contribution to the integral comes from
$|x-X|\sim \sigma$, so that, to leading order in $\sigma/R$, we can
replace the factor $1/x$ by $1/X$ in (\ref{eq:Ifastaa}), and extend the
lower limit of the $x$-integral to $\mp\infty$.  Letting
\begin{equation}
  \label{eq:P_y}
  \Ptrans(y) = \int_{-\infty}^\infty \rmd x\, \left|\psi(x,y)\right|^2
\end{equation}
denote the wavepacket's probability density in $y$ (the direction transverse
to the incident velocity), we obtain,
to leading order in
$\sigma/R$ and $R/(k\sigma^2)$, the expression
\begin{equation}
  \label{eq:Ifast}
\Iexpect = 
\pm 2\rmi\hbar \rme^{\mp\rmi\pi\fracalph}\sin\pi\fracalph\, 
 \Theta(\pm X)\Ptrans(0)
\end{equation}
for the impulse on fast wavepackets.  Here $\Theta(x)$ is the unit step function.

The impulse (\ref{eq:Ifast}) is independent of the wavenumber $k$.  
To leading order, it vanishes for wavepackets which miss the flux
line (eg, $|Y| >> \sigma$, or $\pm X > 0$), as shown previously by Olariu
\& Popescu (1983, 1985).  For fast wavepackets which hit the flux line,
taking $\Ptrans(Y)$ to be of order $1/\sigma$ for $|Y| \sim \sigma$, we
get that the impulse is of order $\hbar/\sigma$. Therefore, it is of
the same order as the dispersion in momentum, in contrast with the case of 
slow wavepackets, for which the impulse is much smaller (by a factor
of $\sigma/R$) than the dispersion.

The $y$-component of the forward impulse, ie the imaginary part of~\eqref{eq:Ifast}, is given 
in the forward case by
\begin{equation}
  \label{eq:Iy}
  \me{\psi}{I_{+y}}{\psi} = \pm \hbar \sin 2\pi\fracalph\, \Theta(\pm X)\Ptrans(y).
\end{equation}
This can be regarded as an analogue in the time domain of Shelankov's
formula (\ref{eq:shelimpulse}) for the transverse momentum imparted to a stationary paraxial beam.

\section{Enclosed and distributed fluxes}\label{sec:encl-distr-flux}

Two well-known regularizations of the Aharonov-Bohm flux line are to
enclose the flux in an impenetrable cylindrical barrier, or to
distribute the flux uniformly in a cylindrical tube.  Here we show
that the force and impulse operators in both cases approach the
Aharonov-Bohm limit, in a sense to be explained, as the radius
$\epsilon$ of the cylinder approaches zero.

In a circularly symmetric gauge, the vector potential for both models is of the form
$\Avec^\epsilon(\rvec) = A^\epsilon(r)\phihat$.  The kinetic momentum
is given by
\begin{equation}
  \label{eq:km2}
M\Vcal^\epsilon = M(V_x^\epsilon + \rmi V_y^\epsilon) =
\frac{\hbar}{\rmi} e^{\rmi\phi}
\left( \partial_r
    +  \frac{\rmi\partial_\phi}{r} + \frac{2\pi}{\Phi_0}A^\epsilon(r)\right),
\end{equation}
and the regularized Hamiltonian by
\begin{equation} \label{eq:Ha}
H^\epsilon = \case12M\big((V^\epsilon_x)^2 + (V^\epsilon_y)^2\big) =
-\frac{\hbar^2}{2M}
\left(\partial^2_r + \frac{\partial_r}{r} + \left(\frac{
      (\rmi\partial_\phi}{r} + 
\frac{2\pi}{\Phi_0}A^\epsilon(r)\right)^2\right).
\end{equation} 
The eigenfunctions of the Hamiltonian and kinetic angular momentum,
with energy $E = \hbar^2k^2/2M$ and kinetic angular momentum $m\hbar$,
are of the form
\begin{equation}
\label{eq:psieps}
\chi^\epsilon_{k,m}(\rvec) = R^\epsilon_{k,m}(r) \rme^{\rmi m \phi}.  
\end{equation}
The radial eigenfunctions $R^\epsilon_{k,m}(r)$ are taken to be real
and normalized, like the Aharonov-Bohm radial eigenfunctions $J_{|m-\alpha|}(kr)$,  according to
\begin{equation}
\label{eq:norm}
\int_0^\infty R^\epsilon_{p,m}(r) R^\epsilon_{k,m}(r) r\,\rmd r = \frac{\delta(k - p)}{k}.
\end{equation}
These conditions determine the radial eigenfunctions up to an overall
sign, which is fixed by requiring that $R^{\epsilon}_{k,m}(r)$
approach $J_{|m - \alpha|}(kr)$ as $\epsilon$ approaches zero.

Let $c_m(k)$ denote a smooth, normalized probability amplitude for energy and
angular momentum.  Let $\psi(\rvec)$ and $\psi^\epsilon(\rvec)$ denote
the corresponding coordinate wavefunctions for the Aharonov-Bohm and
regularized Hamiltonians, respectively.  That is,
\begin{eqnarray}
  \label{eq:corresponding}
  \psi(\rvec) &=& \frac{1}{2\pi}\sum_{m=-\infty}^\infty \int_0^\infty c_m(k)
  J_{|m-\alpha|}(kr)\rme^{\rmi m \phi} k \rmd k,\\
 \psi^\epsilon(\rvec) &=& \frac{1}{2\pi}\sum_{m=-\infty}^\infty \int_0^\infty c_m(k)
 R^\epsilon_{k,m}(r)\rme^{\rmi m \phi} k \rmd k.
\end{eqnarray}
From the preceding discussion, it is clear that 
$\psi^\epsilon(\rvec)$ approaches $\psi(\rvec)$ pointwise as $\epsilon$ approaches
zero.  It turns out that their force and impulse expectation values
also coincide as $\epsilon\rightarrow 0$, ie
\begin{equation}
  \label{eq:forcelimit}
  \lim_{\epsilon\rightarrow0}\me{\psi^\epsilon}{\Fcal^\epsilon}{\psi^\epsilon} =
   \me{\psi}{\Fcal}{\psi},
\end{equation}
\begin{equation}
\label{eq:impulselimit}
 \lim_{\epsilon\rightarrow0} \me{\psi^\epsilon}{\Icalpm^\epsilon}{\psi^\epsilon}
= \me{\psi}{\Icalpm}{\psi}.
\end{equation}
Note that (\ref{eq:forcelimit}) and~\eqref{eq:impulselimit} do not imply, nor is it the case,
that the operators
$\Fcal^\epsilon$ and $\Icalpm^\epsilon$ approach their 
Aharonov-Bohm counterparts, $\Fcal$ and $\Icalpm$, as
$\epsilon$ approaches 0.  Indeed, neither does the regularized
Hamiltonian $H^\epsilon$ approach the Aharonov-Bohm Hamiltonian $H$; given
$\epsilon > 0$, one can construct wavefunctions whose energy
expectation values with respect to $H^\epsilon$ and $H$ differ by
arbitrarily large amounts.

Instead of~\eqref{eq:forcelimit} and (\ref{eq:impulselimit}), we show below, for the enclosed and
distributed fluxes separately, that the eigenstate matrix elements of
the regularized force operator approach the Aharonov-Bohm limit as
$\epsilon\rightarrow 0$, ie,
\begin{equation}
  \label{eq:matrixeps}
   \lim_{\epsilon\rightarrow0}
   \me{\chi^\epsilon_{p,n}}{\Fcal^\epsilon}{\chi^\epsilon_{k,m}}
=  \me{\chi_{p,n}}{\Fcal}{\chi_{k,m}} =
   \frac{2\hbar^2}{M}\sin\pi\fracalph k^\fracalph p^{1-\fracalph}\delta_{m,a}\delta_{n,a+1}.
\end{equation}
Formally, of course,~\eqref{eq:forcelimit} and~\eqref{eq:matrixeps}
are equivalent.  However, for the sake of brevity we shall omit the
details required for a rigorous demonstration.
These details are straightforward to supply, and are similar to those
given in the first part of the Appendix.

The result (\ref{eq:impulselimit}) for the impulse follows from the
corresponding result~\eqref{eq:forcelimit} for the force, once it has
been established that the force expectation values
$\me{\psi^\epsilon}{\Fcal^\epsilon(t)}{\psi^\epsilon}$ and
$\me{\psi}{\Fcal(t)}{\psi}$ are integrable in time.  In the Appendix
it is shown that, in the Aharonov-Bohm case, the force expectation value decays 
as $1/t^2$; a similar argument may be given for the
regularized force.

\subsection{Enclosed flux.}
The kinetic momentum $M\Vcal^\epsilon$ and Hamiltonian $H^\epsilon$
have the same operational form as in the Aharonov-Bohm case, but act
on wavefunctions defined on $r \ge \epsilon$ which vanish at $r =
\epsilon$.  For two such wavefunctions, $\psi^\epsilon(\rvec)$ and $\eta^\epsilon(\rvec)$, 
assumed to be
smooth and normalized, we have
\begin{eqnarray}
  \label{eq:fenclosed} 
\me{\psi^\epsilon}{\Fcal^\epsilon}{\eta^\epsilon} &=&
 \frac{\rmd}{\rmd t}
 \braket{\psi^\epsilon}{M\Vcal^\epsilon\eta^\epsilon} = 
\frac{i}{\hbar} \left[\braket{H^\epsilon
    \psi^\epsilon}{M\Vcal^\epsilon \eta^\epsilon} -
  \braket{\psi^\epsilon}{M\Vcal^\epsilon(H^\epsilon \eta^\epsilon)} \right]
\nonumber\\
&=&\frac{i}{\hbar}\Big( \braket{H^\epsilon
      \psi^\epsilon}{M\Vcal^\epsilon \eta^\epsilon} -\braket{
      \psi^\epsilon}{H^\epsilon\left(M\Vcal^\epsilon \eta^\epsilon\right)} 
\Big).
\end{eqnarray}
The last equality follows from the fact that the commutator
$[H^\epsilon, M\Vcal^\epsilon]$ is proportional to the Lorentz
force operator~\eqref{eq:1}, which vanishes for the
enclosed flux.  However, the final expression in (\ref{eq:fenclosed})
does not vanish; the relation $\braket{H^\epsilon
  \psi^\epsilon}{\xi^\epsilon} = \braket{
  \psi^\epsilon}{H^\epsilon\xi^\epsilon}$, where  
\begin{equation}
\label{eq:xi(r)}
\xi^\epsilon(\rvec) = (M\Vcal^\epsilon \eta^\epsilon)(\rvec), 
\end{equation}
need not hold, 
because $\xi^\epsilon(\rvec)$ 
need not vanish at $r = \epsilon$ (alternatively, $\ket{\xi^\epsilon}$ is not in the
domain 
of $H^\epsilon$).  Indeed, integration by parts in~\eqref{eq:Ha} gives
\begin{eqnarray}
\label{eq:boundaryterm}
\fl \braket{H^\epsilon \psi^\epsilon}{\xi^\epsilon} - \braket{\psi^\epsilon}{H^\epsilon\xi^\epsilon}
&=& 
-\frac{\hbar^2}{2M}\int_0^{2\pi}\!\int_\epsilon^\infty
\left[
\left(
\psi^{\epsilon*}_{rr} + \frac{\psi^{\epsilon*}_r}{r} + (
\frac{\rmi\psi^{\epsilon*}_\phi + \alpha \psi^{\epsilon*} )}{r^2}
\right)\right.
\xi^\epsilon\nonumber\\
&&\left. \qquad -
\psi^{\epsilon*}
\left(
\xi^\epsilon_{rr} + \frac{\xi^\epsilon_r}{r} + \frac{(\rmi \xi^\epsilon_\phi + \alpha \xi^\epsilon)^2}{r^2}
\right)
\right]
r\,\rmd r\,\rmd \phi\nonumber\\
&=& -\frac{\hbar^2}{2M}\int_0^{2\pi}
\psi^{\epsilon*}_r(\epsilon,\phi)\xi^\epsilon(\epsilon,\phi)\,\rmd
\phi
\end{eqnarray}
%  {\psi^\epsilon}^*(M\Vcal^\epsilon H^\epsilon \psi^\epsilon)\right)\, r\rmd r \rmd \phi.
From~\eqref{eq:km2} and~\eqref{eq:xi(r)},
\begin{equation}
  \label{eq:xi}
  \xi^\epsilon(\epsilon,\phi) = \frac{\hbar}{\rmi} \rme^{\rmi\phi} \eta_r(\epsilon,\phi).  
\end{equation}
Substituting~this result into~\eqref{eq:boundaryterm}, we get 
\begin{equation}
\label{eq:Fa2}
\me{\psi^\epsilon}{\Fcal^\epsilon}{\eta^\epsilon} = 
\frac{\hbar^2}{2M} \epsilon
\int_0^{2\pi} \psi_r^{\epsilon*}(\epsilon,\phi)\eta_r^{\epsilon*}(\epsilon,\phi)
\e^{\rmi\phi}\,
\rmd \phi,
\end{equation}
a result obtained previously by Peshkin (1981, 1989).  Note
that if we were to substitute, for $\psi^\epsilon(r)$
and $\eta^\epsilon(r)$, the leading-order behaviour~\eqref{eq:r=0} of Aharonov-Bohm wavefunctions, we
would recover, formally, the Aharonov-Bohm result~\eqref{eq:fexpect} for the force expectation value.

Instead, we take $\psi^\epsilon(\rvec)$ and $\eta^\epsilon(\rvec)$ in~\eqref{eq:Fa2} to be 
eigenfunctions of the regularized Hamiltonian.  Then
\begin{equation}
\label{eq:Fa3}
\me{\chi^\epsilon_{p,n}}{\Fcal^\epsilon}{\chi^\epsilon_{k,m}} = 
\frac{\pi \hbar^2}{M} \epsilon {R^\epsilon_{p,m+1}}'(\epsilon)
{R^\epsilon_{k,m}}'(\epsilon) \delta_{n,m+1}.
\end{equation}
(Strictly speaking, this is not legitimate, as there would appear
boundary terms at $r = \infty$ in~\eqref{eq:boundaryterm}.  However,
these would vanish when we consider expectation values, as in (\ref{eq:forcelimit}).)

To evaluate (\ref{eq:Fa3}) we need the derivatives of the radial
eigenfunctions at $r = \epsilon$.
The radial wavefunctions themselves are given by
\begin{equation}
\label{eq:radial}
R^\epsilon_{k,m}(r) = {C^\epsilon_{k,m}}\left(N_{|m-\alpha|}(k\epsilon)J_{|m-\alpha|}(kr) -
J_{|m-\alpha|}(k\epsilon)N_{|m-\alpha|}(kr)\right),
\end{equation}  
where $N_\nu(z)$ is the Neumann function.   The constant
$C^\epsilon_{k,m}$ is determined by the normalization condition
(\ref{eq:norm}), and is given by
\begin{equation}
\label{eq:normfact}
C^\epsilon_{k,m} = \left(J_{|m-\alpha|}^2(k\epsilon) + N^2_{|m-\alpha|}(k\epsilon)\right)^{-\case12},
\end{equation}
and, to leading order in $\epsilon$, by
\begin{equation}
  \label{eq:Ceps}
  C^\epsilon_{k,m} = \left|N_{m-\alpha}(k\epsilon)\right|^{-1} 
  =  \frac{\pi}{\Gamma(|m-\alpha|)}
\left(\frac{k\epsilon}{2}\right)^{|m-\alpha|}.
\end{equation}
The Wronskian relation,
$J_\nu(z) N'_\nu(z)- J'_\nu(z) N_\nu(z) = 2/(\pi z)$, implies that
\begin{equation}
  \label{eq:R'}
  {R^\epsilon_{k,m}}'(\epsilon) = -\frac{1}{\pi (\epsilon/2)} C^\epsilon_{k,m},\quad
{R^\epsilon_{p,m+1}}'(\epsilon) = -\frac{1}{\pi (\epsilon/2)} C^\epsilon_{p,m+1}.
\end{equation}

Substituting  (\ref{eq:Ceps}) and (\ref{eq:R'}) into (\ref{eq:Fa3}), we obtain, to leading order in $\epsilon$,
\begin{equation}
  \label{eq:Fpenult}
\fl \me{\chi^\epsilon_{p,n}}{\Fcal^\epsilon}{\chi^\epsilon_{k,m}} = 
\frac{2\pi\hbar^2}{M}\frac{k^{|m-\alpha|}
p^{|m+1-\alpha|}}{\Gamma(|m+1-\alpha|)\Gamma(|m-\alpha|)}
\left(\frac{\epsilon}{2}\right)^{|m+1-\alpha|+|m-\alpha| - 1}\delta_{m,n+1}.
\end{equation}
In the limit $\epsilon\rightarrow 0$, only the $m = a$ term survives, and the reflection formula
for the $\Gamma$-function gives
\begin{eqnarray}
  \label{eq:Flast}
 \lim_{\epsilon\rightarrow 0} \me{\chi^\epsilon_{p,n}}{\Fcal^\epsilon}{\chi^\epsilon_{k,m}} &=& 
\frac{2\pi\hbar^2}{M}\frac{k^{\fracalph}
p^{1-\fracalph}}{\Gamma(1-\fracalph)\Gamma(\fracalph)}
\delta_{m,a}\delta_{n,a+1}\nonumber\\
&=& \frac{2\hbar^2}{M}\sin\pi\fracalph \,k^\fracalph
  p^{1-\fracalph}\delta_{m,a} \delta_{n,a+1},
\end{eqnarray}
in accord with (\ref{eq:matrixeps}).

%The corresponding limit for the impulse, (\ref{eq:corlimitim}), holds
%if the force $\me{\psi^\epsilon(t)}{\Fcal^\epsilon}{\psi^\epsilon(t)}$
%is integrable in time.  This can be established from~\eqref{eq:Fa2},
%and the fact that $\psi^\epsilon(\epsilon,\phi,t)$ falls off at least
%as fast as $1/t$ (which follows from the conditions~(\ref{eq:prop0})
%upon integrating by parts with respect to $k$ in the
%expansion~\eqref{eq:wpa}).  **Need to elaborate on this.**

\subsection{Distributed flux}
The distributed flux model was used by Nielsen and Hedeg\aa rd
(1995) to obtain, from the force balance equations, the on-shell matrix
elements of the force in the limit $\epsilon\rightarrow 0$. Here we
carry out a different calculation to obtain the general matrix elements of
the force.

It suffices to consider the case $\alpha > 0$ (the case of negative
flux is obtained from time-reversal).  The vector potential is given
by
\begin{eqnarray}\label{eq:Aeps}
A^\epsilon(r) &= \alpha  \Phi_0 r/ (2\pi \epsilon^2), & \quad r < \epsilon,\\
       &= \alpha \Phi_0 / 2\pi r, & \quad r\ge \epsilon,
\end{eqnarray}
corresponding to the magnetic field $B^\epsilon(r) = (\alpha \Phi_0/\pi
\epsilon^2) \Theta(\epsilon - r)$, where $\Theta(x)$ is the unit step function.  In
this case, the force operator is just  the Lorentz force (\ref{eq:1}).  It
is convenient to introduce the dimensionless radial coordinate $u =
r^2/\epsilon^2$, so that the interior of the flux tube is given by
$0\le u \le 1$.  The kinetic momentum is given by 
\begin{equation}
\label{eq:MVu}
M\Vcal^\epsilon = 
\frac{\hbar}{\rmi} \rme^{\rmi \phi} \frac{u^{1/2}}{\epsilon}\left(2\partial_u + 
\frac{\rmi\partial_\phi}{u}+\alpha\right)
\end{equation}
Then 
\begin{eqnarray}
  \label{eq:Lordist}
  \Fcal &=& -\rmi \frac{e}{2Mc}\left(M\Vcal^\epsilon B^\epsilon + B^\epsilon
  M\Vcal^\epsilon\right)\nonumber\\
& =& -\frac{2\hbar^2}{M\epsilon^3}\alpha\rme^{\rmi\phi} u^{\case12}
\left[\Theta(1-u)\left(2\partial_u + \rmi \frac{\partial_\phi}{u} + \alpha\right) - \delta(u-1)\right].
\end{eqnarray}
The matrix elements of the force are given by
\begin{eqnarray}
\label{eq:fdist}
\fl \me{\chi^\epsilon_{p,n}}{\Fcal^\epsilon}{\chi^\epsilon_{k,m}} = 
      - 2\pi\frac{ \hbar^2}{M\epsilon} \alpha  \delta_{n,m+1}\nonumber\\
\times \left[
        \int_0^1 
           T^\epsilon_{p,m+1} \left(2{T^\epsilon_{k,m}}' + \left(\alpha -
               \frac{m}{u} \right)T^\epsilon_{k,m}\right) u^{\case12}\,\rmd u -
          t^\epsilon_{p,m+1}t^\epsilon_{k,m}\right],
\end{eqnarray}
where $T^\epsilon_{k,m}(u)$ denotes the 
radial eigenfunction expressed in terms of the scaled variable $u$,
and $t^\epsilon_{k,m} = T^\epsilon_{k,m}(1).$

Inside the flux tube, the radial eigenfunctions are given by (Landau \& Lifschitz 1965)
\begin{equation}
\label{eq:Teps}
\fl T^\epsilon_{k,m}(u) = C^\epsilon_{k,m} \e^{-\alpha u/2} u^{|m|/2} 
M\left(-\frac{(k\epsilon)^2}{4\alpha} + \frac{|m|-m+1}{2},
  |m|+1,\alpha u\right),\ \ 0 \le u \le 1,
\end{equation}
where $M(a,b,z)$ is the confluent hypergeometric function (Abramowitz
\& Stegun 1970).  Outside the flux tube,
\begin{equation}
  \label{eq:Tepsout}
  R^\epsilon_{k,m}(r) = D^\epsilon_{k,m} J_{|m-\alpha|}(kr) +
  E^\epsilon_{k,m} N_{|m-\alpha|}(kr),\ \  r \ge \epsilon.
\end{equation}
The coefficients $C^\epsilon_{k,m}$, $D^\epsilon_{k,m}$ and
$E^\epsilon_{k,m}$ are determined by requiring the radial
eigenfunction and its first derivative to be continuous at $r =
\epsilon$ (the second derivative is then continuous there as well, as it
turns out), and by the normalization condition
\begin{equation}
  \label{eq:norm3}
  (D^\epsilon_{k,m})^2 + (E^\epsilon_{k,m})^2 = 1,
\end{equation}
which follows from (\ref{eq:norm}).  

To evaluate the force matrix element (\ref{eq:fdist}), we
only require the function inside the flux cylinder.  Straightforward
algebra gives the coefficient $C^\epsilon_{k,m}$, to leading order in
$\epsilon$, as
\begin{equation}
  \label{eq:Cepsapprox}
  C^\epsilon_{k,m} =  
\frac{2\e^{\alpha/2}(\case12 k\epsilon)^{|m-\alpha|}}
{\Gamma(|m-\alpha|)
\left[(|m-\alpha|+|m|-\alpha)
%M(\case12(|m|+1-m), |m|+1,\alpha) 
f_m
+2
%\alpha M'(\case12(|m| + 1 - m), |m|+1,\alpha)
f'_m
\right]},
\end{equation}
where
\begin{equation}
F_m(u) = M(\case12(|m|-m+1), |m|+1,\alpha u),
\end{equation}
and $f_m$ and $f'_m$ denote the values of $F_m$ and $F'_m$ at $u = 1$.

Substituting (\ref{eq:Teps}) and (\ref{eq:Cepsapprox}) into (\ref{eq:fdist}), we find that 
that $\me{\chi^\epsilon_{p,n}}{\Fcal^\epsilon}{\chi^\epsilon_{k,m}}$
is of order $\epsilon^{|m-\alpha| + |m+1-\alpha| - 1}$, and therefore
vanishes in the limit $\epsilon\rightarrow 0$ unless $m = a$.
We obtain
\begin{equation}
  \label{eq:fdistpen}
\lim_{\epsilon\rightarrow 0}
\me{\chi^\epsilon_{p,n}}{\Fcal^\epsilon}{\chi^\epsilon_{k,m}} = 
\frac{2\hbar^2}{M}\sin \pi\fracalph\,
 k^{\fracalph}p^{1-\fracalph} \frac{L(\alpha)}{R(\alpha)} \delta_{m,a}\delta_{n,a+1},
\end{equation}
where
\begin{eqnarray}
\label{eq:LR}
L(\alpha) &=&   -2\alpha \int_0^1 F_a' F_{a+1}
\rme^{\alpha(1-u)}u^{a+1}\,\rmd u + \alpha f_a f_{a+1},\nonumber\\
R(\alpha) &=& 2 (p(1)f_{a+1} + f'_{a+1}f'_a),
\end{eqnarray}
and
\begin{equation}
\label{eq:p(u)}
p(u) = a + 1 - \alpha u.
\end{equation}
As we show below, $L(\alpha) = R(\alpha)$, or, equivalently,
\begin{equation}
  \label{eq:identity}
 \int_0^1  2\alpha F_a' F_{a+1} \rme^{\alpha(1-u)}u^{a+1}\,\rmd u =  \alpha f_a f_{a+1}-
  2(p(1)f_{a+1} + f_{a+1}')f_a'.
\end{equation}
With this identity,~\eqref{eq:fdistpen} gives the required result (\ref{eq:matrixeps}).

To establish the identity~\eqref{eq:identity}, it is convenient to express $F_{a+1}$ in terms
of $F_a$ by means of the recurrence relation (Abramowitz \& Stegun 1970)
\begin{equation}
  \label{eq:Mrecur}
  (a+\case12) M(\case12, a+2, u) = (a+1)\left(M(\case12, a + 1, u) - M'(\case12,a+1,u)\right),
\end{equation}
which implies that 
\begin{equation}
\label{eq:Frecur}
(a+\case12)F_{a+1} = (a+1)(F_a - F_a'/\alpha).  
\end{equation}
With the
differential equation
\begin{equation}
  \label{eq:diffeq}
  u F_a'' = -p(u) F_a' + \frac{\alpha}{2}F_a,
\end{equation}
it is straightforward to show that the integrand on the left-hand side
of (\ref{eq:identity}) is given by $W'(u)$, where
\begin{equation}
  \label{eq:W(u)}
  W(u) = 2\frac{a+1}{a+\frac{1}{2}} u^{a+1} e^{\alpha(1-u)}
\left( \frac{\alpha}{2} F_a^2 - p(u)F_aF_a' - u{F_a'}^2
\right).
\end{equation}
$W(0)$ vanishes, whereas $W(1)$, with the aid of~\eqref{eq:Frecur} and~\eqref{eq:diffeq}, is seen to be equal to 
the right-hand side of~\eqref{eq:identity}.

%\section{Summary}\label{sec:conclusion}
%We give a brief summary.  Using kinematic arguments, in particular modified commutation relations consistent
%with gauge invariance, we obtained the Lorentz force~\eqref{eq:forceop} due to an Aharonov-Bohm flux line.
%The stationary state matrix elements,~\eqref{eq:matrix}, agree with previously obtained expressions when the
%states have the same energy.  Expectation values for a stationary beam yielded an exact version~\eqref{eq:stationary}
%of Shelankov's formula~\eqref{eq:shelforce} (Shalankov 1997, Shelankov 2000).  Integration in time yielded the
%impulse operator,~\eqref{eq:impulse}, in the position representation.  We computed expectation 
\ack
We thank Professor Sir MV Berry and Professor D Khemelnitskii for  stimulating discussions.
\appendix
\section{Wavepacket expectation values}
\label{sec:rigo-deriv-ahar}
The force and impulse due to an Aharonov-Bohm flux line can be
calculated rigorously for suitably well-behaved wavefunctions
$\psi(\rvec)$.  We will take these to be such that
\begin{equation}
\label{eq:prop0}
\fl \qquad \parbox{5.25in}{\rm $c_m(k) = \braket{\chi_{k,m}}{\psi}$ is
  smooth in $k$ and falls off, along
with its derivatives, faster than any power of $k$ and $m$.}
\end{equation}  
Using standard arguments, one can show that~\eqref{eq:prop0} implies the following properties of $\psi(\rvec)$
and $(H\psi)(\rvec)$, where $H$ is the Aharonov-Bohm Hamiltonian:
\begin{equation}
\label{eq:properties}
\fl \qquad \parbox{5.25in}{\rm $\psi(\rvec)$ and $(H\psi)(\rvec)$ are smooth for $r > 0$ and
fall off, along with their derivatives, faster than any
power of $r$,\nonumber} 
\end{equation}
and
\begin{eqnarray}
  \label{eq:prop2}
  \psi_m(r)&=& C_m r^{|m-\alpha|} + \Or\left(r^{|m-\alpha| + 1}\right),\nonumber\\
  (H\psi)_m(r)&=&D_m r^{|m-\alpha|} + \Or\left(r^{|m-\alpha| + 1}\right),
\end{eqnarray}
where, in general, 
\begin{equation}
\eta_m(r) = \frac{1}{2\pi}\int_0^{2\pi} \eta(r,\phi) \rme^{-\rmi m
\phi}\,\rmd \phi.  
\end{equation}
(In fact, properties
(\ref{eq:properties}) and (\ref{eq:prop2}) are also shared by
$(H^j\psi)(\rvec)$, for $j > 1$.  The argument to follow would hold
under weaker conditions,
but we assume (\ref{eq:prop0}) for simplicity.)

The expectation value of force is given by
\begin{eqnarray}
  \label{eq:frig1}
  \me{\psi}{\Fcal}{\psi} &=& \int\!\int\left({\dot \psi}^* (M\Vcal\psi) + \psi^* (M\Vcal\dot\psi)\right)\,\rmd^2r\nonumber\\
&=& \frac{1}{\rmi \hbar}\int\!\int\Big(-({H\psi}^*) (M\Vcal\psi) + \psi^* (M\Vcal H\psi)\Big)\,\rmd^2r,
\end{eqnarray}
where $M\Vcal$ is given by~\eqref{eq:13}.
From (\ref{eq:properties}) and (\ref{eq:prop2}), it is evident that the
$\rvec$-integral in (\ref{eq:frig1}) converges absolutely.  This
allows us to introduce a factor $\exp(-\epsilon^2 r^2)$ in the
integrand, and then take the limit of the integral as $\epsilon\rightarrow
0$.  This Gaussian factor will justify subsequent reorderings of
operations.  Note that the integral cannot be
expressed in terms of the expectation value of the commutator
$[H,M\Vcal]$, because of the singularity in the radial derivative of $\psi(\rvec)$
at the origin
(specifically, 
$(M\Vcal\psi)(\rvec)$ is not in the domain of $H$).

We introduce the eigenfunction expansion
\begin{equation}
  \label{eq:psiexpand}
  \psi(\rvec) = \frac{1}{2\pi}\sum_{m=-\infty}^{\infty} \int_0^\infty
  c_m(k) J_{|m-\alpha|}(kr) \rme^{\rmi m \phi} k \rmd k,
\end{equation}
and a similar expansion for $(H\psi)(\rvec)$, with  $c_m(k)$
replaced by $-(\hbar^2 k^2/2M) c_m(k)$.  Using standard arguments, one
can show that (\ref{eq:prop0}) implies that the differential operator $M\Vcal$, when applied to
$\psi$ and $H\psi$, can be taken inside the $m$-sum and $k$-integral.
The recurrence relation,
\begin{equation}
  \label{eq:recurrence}
  J_{\nu \pm 1}(z) = \mp \left(J_\nu'(z) \mp \frac{\nu}{z} J_{\nu}(z)\right),
\end{equation}
implies that
\begin{eqnarray}
  \label{eq:MVbessel}
\fl  M\Vcal \left(J_{|m-\alpha|}(kr)\rme^{\rmi m\phi}\right) &= \sgn(m-a)\rmi\hbar k
%  \chi_{k,m+1}(\rvec)
J_{|m+1-\alpha|}(kr)\rme^{\rmi (m+1)\phi}, &\quad m\ne a,\nonumber\\
  &= -\rmi \hbar k J_{\fracalph - 1}(kr) \rme^{\rmi (a+1)\phi}, &\quad m
  = a.
\end{eqnarray}
Substituting (\ref{eq:psiexpand}) and (\ref{eq:MVbessel}) into
(\ref{eq:frig1}), along with the eigenfunction expansion of $\psi^*(\rvec)$ with
coefficients $c_n^*(p)$, we obtain
% Need tex manual to do this
\begin{eqnarray}
  \label{eq:frig2}
 \fl\me{\psi}{\Fcal}{\psi} = \lim_{\epsilon\rightarrow 0}
 \frac{\hbar^2}{8\pi^2M}\int_0^\infty  \rme^{-\epsilon^2 r^2}r\rmd r\int_0^{2\pi}\rmd \phi\,
\sum_{m = -\infty}^\infty \sum_{n = -\infty}^\infty 
 \rme^{\rmi(m + 1 - n)\phi} \int_0^\infty \rmd p \int_0^\infty \rmd
 k\nonumber\\
 \fl \qquad  \times c_n^*(p) c_m(k)
 k^2 p (k^2 - p^2)J_{|n-\alpha|}(pr)
\times \cases{\sgn(m-a) J_{|m+1-\alpha|}(kr),& $m\ne a$,\\
        -J_{\fracalph - 1}(kr),&$m=a$.\\}
\end{eqnarray}

The sums and
integrals in (\ref{eq:frig2}) are uniformly and absolutely convergent,
and can be interchanged. On performing the $\phi$-integral, the
sum on $n$  collapses to the single term $n = m+1$. We obtain
\begin{equation}
  \label{eq:frig3}
  \me{\psi}{\Fcal}{\psi} = \frac{\hbar^2}{4\pi M} \lim_{\epsilon\rightarrow 0}
  \sum_{m=-\infty}^\infty \int_0^\infty \int_0^\infty
  K^\epsilon_m(k,p) c_{m+1}^*(p) c_m(k)\,\rmd k \rmd p,
\end{equation}
where, for $m\ne a$,
\begin{equation}
\label{eq:Keps2mnea}
\fl K^\epsilon_m(k,p) = \sgn(m-a) k^2 p (k^2 - p^2) 
\int_0^\infty \rme^{-\epsilon^2
  r^2}J_\nu(pr) J_\nu(kr) r \,\rmd r, \ \ \nu = |m-\alpha + 1|,
\end{equation}
and, for $m = a$,
\begin{equation}
\label{eq:Keps2mea}
K^\epsilon_a(k,p) =  k^2 p (p^2 - k^2) \int_0^\infty  \rme^{-\epsilon^2 r^2}
  J_\nu(pr) J_{-\nu}(kr) r \,\rmd r, \quad \nu = 1-\fracalph.
\end{equation}
Below, in \ref{secmnea}, we show that the
contributions from the $m\ne a$ terms vanish in the limit, ie
\begin{equation}
  \label{eq:mnea}
  \lim_{\epsilon\rightarrow 0} \sum_{m\ne a} \int_0^\infty\!
  \int_0^\infty K^{\epsilon}_m(k,p)  c_{a+1}^*(p) c_a(k)\,\rmd k \rmd
  p = 0,
\end{equation}
while, in \ref{secmea}, we show for the $m = a$ term that
\begin{eqnarray}
\label{eq:mea}
 \fl \lim_{\epsilon\rightarrow 0} \int_0^\infty\!
  \int_0^\infty K^\epsilon_a(k,p) c_{a+1}^*(p) c_a(k)\,\rmd k \rmd p \nonumber\\
\lo=
  \frac{2}{\pi}\sin\pi \fracalph \int_0^\infty
  \int_0^\infty
  k^{1+\fracalph}p^{2-\fracalph} c_{a+1}^*(p) c_a(k)\,\rmd k \rmd p.
\end{eqnarray}
Substitution of~\eqref{eq:mnea} and~\eqref{eq:mea} into (\ref{eq:frig3}) gives
\begin{equation}
  \label{eq:frig4}
  \me{\psi}{\Fcal}{\psi} = \frac{\hbar^2}{2\pi^2 M}\sin\pi \fracalph 
\int_0^\infty \int_0^\infty
   k^{1+\fracalph}p^{2-\fracalph}  c_{a+1}^*(p) c_a(k)\,\rmd k \rmd p.
\end{equation}
This is equivalent to the matrix element (\ref{eq:matrix}), obtained formally
in Section~\ref{sec:force-operator}.

We note that, with $\epsilon = 0$, the integrals in (\ref{eq:Keps2mnea}) and 
(\ref{eq:Keps2mea})
correspond to singular (ie, not absolutely convergent) cases of the
discontinuous Weber-Schafheitlin integral (Abramowitz \& Stegun 1970).
Formal evaluation of these integrals would give
(\ref{eq:mnea}) and (\ref{eq:mea}) immediately.  The arguments
in~\ref{secmnea} and \ref{secmea} serve to justify these formal results.

To obtain 
the force expectation value~\eqref{eq:fexpect}, we
express $c_a(k)$ and $c^*_{a+1}(p)$ in (\ref{eq:frig4}) in terms of
$\psi(\rvec)$ to obtain
\begin{eqnarray}
  \label{eq:frig5}
\fl   \me{\psi}{\Fcal}{\psi} = \frac{2 \hbar^2}{M}\sin\pi \fracalph
\lim_{\epsilon\rightarrow 0} 
\int_0^\infty \int_0^\infty
   k^{1+\fracalph}p^{2-\fracalph}\rme^{-\epsilon^2 (k^2+p^2)} \nonumber\\ \times \left(\int \psi_{a+1}^*(s)
     J_{1-\fracalph}(ps)s\,d s\right)
 \left(\int \psi_a(r)
     J_{\fracalph}(kr)r\,\rmd r
       \right)
\,\rmd k \rmd p,
\end{eqnarray}
Note that the convergence factor $\exp(-\epsilon^2(k^2 + p^2))$ can be
introduced, and the limit $\epsilon\rightarrow 0$ taken outside the integral, since, by
(\ref{eq:prop0}), the $k$- and $p$-integrals in (\ref{eq:frig4}) are
absolutely convergent.  By (\ref{eq:properties}) and (\ref{eq:prop2}),
the $r$- and $s$-integrals in (\ref{eq:frig5}) are absolutely convergent,
so that, for $\epsilon> 0$, we can interchange the order of integration.
The $k$- and $p$-integrals can be
evaluated using (\ref{eq:integral}), with the result
\begin{eqnarray}
  \label{eq:frig6}
  \me{\psi}{\Fcal}{\psi} &=& \frac{2 \hbar^2}{ M}\sin\pi \fracalph 
\lim_{\epsilon\rightarrow 0} 
    \int_0^\infty 
       \left(\frac{\psi_a(r)}{r^\fracalph}\right)
       \rme^{-r^2/4\epsilon^2} 
       \left(
          \frac{r^2}{4\epsilon^2}
       \right)^{\fracalph} 
    \,\rmd \left(\frac{r^2}{4\epsilon^2}\right) 
\nonumber\\
&&\qquad\times \int_0^\infty \left(\frac{\psi^*_{a+1}(s)}{s^{1-\fracalph}}\right)
  \rme^{-s^2/4\epsilon^2} 
\left(\frac{s^2}{4\epsilon^2}\right)^{1-\fracalph} 
  \,\rmd \left(\frac{s^2}{4\epsilon^2}\right) \nonumber\\
&=& \frac{2 \pi\hbar^2}{ M}\sin\pi \fracalph C_a C_{a+1}^*
\Gamma(1+\fracalph)\Gamma(2-\fracalph)\nonumber\\
&=& \frac{4 \pi\hbar^2}{M}  \fracalph (1-\fracalph) C_a C_{a+1}^*,
\end{eqnarray}
where the coefficients $C_a$ and $C_{a+1}$ are given in~\eqref{eq:prop2}.
This is just the result (\ref{eq:fexpect}) of
Section~\ref{sec:force-operator}.

Concerning the impulse, it is straightforward to justify, using arguments like the preceding ones, 
the calculations of
Section~\ref{sec:impulse-operator} 
leading to \eqref{eq:iexpect2a}.
It is only necessary to check that the time-dependent expectation value,
$\me{\psi(t)}{\Fcal}{\psi(t)}$, is integrable in $t$.  
$\me{\psi(t)}{\Fcal}{\psi(t)}$ is
given by an expression like (\ref{eq:frig4}), but with $c_a(k)$ and $c^*_{a+1}(p)$ modulated
by the factors $\exp (-\rmi\hbar k^2 t/2M)$ and $\exp (\rmi\hbar p^2
t/2M)$ respectively.  We have that
\begin{eqnarray}
  \label{eq:1/t}
\fl  \int_0^\infty   k^{1+\fracalph} c_m(k) \exp\left(-\rmi \frac{\hbar k^2}{2M} t\right)
  \rmd k \nonumber\\
\lo = \frac{\rmi M}{\hbar t}\int_0^\infty  \frac{\rmd}{\rmd k}\left(k^\fracalph
  c_m(k)\right) \exp\left(-\rmi \frac{\hbar k^2}{2M} t\right) \rmd k = 
\Or(1/t)
\end{eqnarray}
for large $t$ (the integration by parts is justified by~\eqref{eq:properties} and~\eqref{eq:prop2}).
Similarly, $\int_0^\infty p^{2-\fracalph}
c_{a+1}^*(p) \exp(\rmi\hbar p^2 t/2M) \,\rmd p = \Or(1/t)$.  It
follows that $\me{\psi(t)}{\Fcal}{\psi(t)}$ falls off as $1/t^2$.

\subsection{Proof of (\ref{eq:mnea})}
\label{secmnea}
Given functions $f$ and $g$ defined on a domain $D$, we will say that
$f$ is dominated by $g$ if, for some constant $C$, $|f| < C g$
throughout $D$.  For functions indexed by an
integer $m$, eg $f_m$ and $g_m$, we will say that
$f_m$ is dominated by $g_m$ if $|f_m| < C g_m$ for some $C$ which does not depend on $m$. 
Thus, from (\ref{eq:prop0}), 
\begin{equation}
\label{eq:dominusa}
\fl \qquad \hbox{$c_m(k)c^*_{m+1}(p)$ is
dominated by $(1 + m^2)^{-1} (1 + k^2 + p^2)^{-4}$ for $k,p > 0$.}
\end{equation}

The integral in (\ref{eq:Keps2mnea}) can be evaluated  (Gradshteyn and
Ryzhik 1980) 
to give
\begin{equation}
  \label{eq:abram1}
   K^\epsilon_m(k,p) = 
   \frac{\sgn(m-a)}{2\epsilon^2}k^2 p 
   (k^2 - p^2)\rme^{-(k^2 + p^2)/4\epsilon^2}I_\nu\left(\frac{kp}{2\epsilon^2}\right),
\end{equation}
where $I_\nu(z)$ is a modified Bessel function, and $\nu =
|m-\alpha|$.  From the 
asymptotic behaviour of $I_\nu(z)$ for large argument, 
it follows that 
$I_\nu(z)$ is dominated by  $\rme^z/ \sqrt{z}$ for $z$ real.   Therefore,
the left-hand side of
(\ref{eq:mnea}) is dominated by 
\begin{equation}
  \label{eq:dominus}
\sum_{m=-\infty}^\infty
\frac{1}{1+ m^2} \frac{1}{\epsilon}\int_0^\infty \int_0^\infty 
  \exp\left(-\frac{(k-p)^2}{2\epsilon^2}\right)\frac{k^{3/2} p^{1/2}|k^2 - p^2|}{(1 + k^2 + p^2)^4}  \,\rmd k\rmd p.
\end{equation}
Let us divide the domain of the $(k,p)$-integral into
regions inside and outside the strip
$|k - p| <
\epsilon^{2/3}$.
Inside, the integrand is dominated by  
$\epsilon^{2/3}k^3/(1 + k^2)^4$; thus the integral over the strip is
dominated by $\epsilon^{4/3}$.  Outside the strip, the integrand is
dominated by $\exp\left(-\epsilon^{-2/3}\right)/(1+k^2 + p^2)^2$; thus the
$(k,p)$-integral outside the strip is dominated by
$\exp\left(-\epsilon^{-2/3}\right)$. Therefore, the expression in
\eqref{eq:dominus}
is dominated by 
\begin{equation}
  \label{eq:dominus4}
  \sum_{m=-\infty}^\infty
\frac{1}{1+ m^2} \frac{1}{\epsilon}\left(\epsilon^{4/3} +
    \exp\left(-\epsilon^{-2/3}\right)\right),
\end{equation}
which vanishes as 
$\epsilon\rightarrow
0$.

\subsection{Proof of (\ref{eq:mea})}
\label{secmea}
Substituting the series expansion
\begin{equation}
\label{besselpower}
  J_\nu(z) = \left(\frac{z}{2}\right)^\nu\sum_{u=0}^\infty \frac{1}{u! \Gamma(u+\nu+1)}\left(-\frac{z^2}{4}\right)^u,
\end{equation}
and a similar expansion for $J_{-\nu}(z)$, we get that
\begin{eqnarray}
  \label{eq:besselint}
 \fl  k^\nu \int_0^\infty \rme^{-\epsilon^2 r^2} J_\nu(pr)
 J_{-\nu}(kr) r\,\rmd r\nonumber\\
 \lo = \frac{1}{2\epsilon^2}p^\nu 
  \sum_{u=0}^\infty \sum_{v=0}^\infty
\frac{\Gamma(u+v+1)}{u! v!
  \Gamma(u-\nu+1)\Gamma(v+\nu+1)}\left(-\frac{k^2}{4\epsilon^2}\right)^u
\left(-\frac{p^2}{4\epsilon^2}\right)^v.
\end{eqnarray}
For $\epsilon > 0$, the $r$-integral and $u-$ and $v$-sums are absolutely
convergent for all $k,p \ge 0$.  Inserting in \eqref{eq:besselint} the integral representation
for the reciprocal of the  beta-function (Gradshteyn and Ryzhik 1980)
\begin{eqnarray}
\fl \frac{\Gamma(u+v + 1)}{\Gamma(u - \nu + 1)\Gamma(v + \nu + 1)} =
  \frac{1}{(u+v+1)B(u-\nu + 1,v+\nu+1)} \nonumber\\
\lo = 
\frac{2}{\pi}\re \int_0^{\pi/2}\left(2\rmi\sin \tau \rme^{-\rmi \tau}\right)^u
 \left(-2\rmi\sin \tau \rme^{\rmi \tau}\right)^v \rme^{2\rmi\nu
  (\tau - \pi/2)}\,\rmd \tau,
\end{eqnarray}
we may 
perform the sums  to obtain
\begin{eqnarray}
\label{eq:besselint2}
\fl k^\nu \int_0^\infty \rme^{-\epsilon^2 r^2} J_\nu(pr) J_{-\nu}(kr) r\,\rmd
  r\nonumber\\
\fl \qquad  =   \frac{p^\nu}{\pi\epsilon^2}\re
\int_0^{\pi/2}\exp\left(-\frac{p^2 + k^2}{2\epsilon^2} \sin^2\tau +
  \rmi\frac{p^2 - k^2}{4\epsilon^2}\sin 2\tau +
  2\rmi\nu(\tau-\pi/2)\right)\,\rmd \tau.
\end{eqnarray}
Substituting this result into (\ref{eq:Keps2mea}), we get 
\begin{equation}
  \label{eq:Kepsm=a}
K^\epsilon_a(k,p) = \frac{1}{\pi\epsilon^2} k^{1+\fracalph}
p^{2-\fracalph} (p^2-k^2) \re \int_0^{\pi/2} \rme^{-S/\epsilon^2}\,\rmd\tau,
\end{equation}
where the exponent $S$ is given by
\begin{equation}
  \label{eq:S}
  S = \case14 (k^2 + p^2)(1 - \cos2\tau)  -
  \case14 \rmi (p^2 - k^2)\sin 2\tau -
   2\rmi\epsilon^2(1-\fracalph)(\tau -\pi/2).
\end{equation}

It is clear that the main contribution to the $\tau$-integral in~\eqref{eq:Kepsm=a} comes from the neighbourhood of $\tau = 0$.
If $S$ is expanded about $\tau = 0$ to second order, the
$\tau$-integral yields an error function, whose
leading-order asymptotics as $\epsilon\rightarrow 0$ leads directly to the
required result~\eqref{eq:mea}.  However, the next term in the asymptotic expansion
is not uniformly bounded in $k$ and $p$ --  it contains a factor
$(k^2-p^2)^{-2}$ -- so we must take some additional care.

To proceed, we divide the domain of the $k,p$-integral into the three regions 
specified below, writing
the left hand side of~\eqref{eq:mea} as 
\begin{equation}
  \label{eq:3parts}
\left(\lim_{\epsilon\rightarrow 0}\int\!\int_{D_1} + \lim_{\epsilon\rightarrow 0}\int\!\int_{D_2}
+ \lim_{\epsilon\rightarrow 0}\int\!\int_{D_3}\right)
K^\epsilon_a(k,p) c_{a+1}^*(p) c_a(k)\,\rmd k \rmd p,
\end{equation}
and analyzing the contribution from each region separately.

Let
$D_1$ denote the region $k,p\ge 0$, $k^2 + p^2 \le \epsilon^\beta$, where $\beta$
is chosen to satisfy
\begin{equation}
  \label{eq:betaineq}
  \case{4}{7}<\beta<\case{2}{3}.
\end{equation}
In this region, the exponential factor in (\ref{eq:Kepsm=a}) is
bounded, and $k^{1+\fracalph}p^{2-\fracalph}(p^2-k^2)$ is dominated by
$\epsilon^{5\beta/2}$.  The coefficients $|c_a(k)|$ and $|c^*_{a+1}(p)|$
are bounded, and the area of $D_1$ is dominated by $\epsilon^{2\beta}$, so that 
the contribution from $D_1$ in~\eqref{eq:3parts} is dominated
by $\epsilon^{7\beta/2 - 2}$. Given the choice of $\beta$, this vanishes as
$\epsilon\rightarrow 0$.

Let $D_2$ be the region $k,p\ge 0$, $k^2 + p^2 \ge \epsilon^\beta$ and
$|p^2 - k^2| \le \epsilon^\gamma$, where $\gamma$ is chosen to satisfy
\begin{equation}
  \label{eq:gammaineq}
  \case12+\case14 \beta < \gamma < 1 - \case12 \beta.
\end{equation}
(since $\beta < \case23$, the inequality~\eqref{eq:gammaineq} can be always be satisfied).
Since $1 - \cos2\tau \ge  \tau^2$ for $0\le \tau \le \pi/2$,
the factor $\exp(-S/\epsilon^2)$ is dominated by the Gaussian $\exp(-\sigma^2 
 \tau^2)$, where $\sigma = \case12\epsilon^{\beta/2 - 1}$.  Thus $\int_0^{\pi/2} \exp(-S/\epsilon^2)\,\rmd \tau$
is dominated by $\epsilon^{1-\beta/2}$.  Then, from~\eqref{eq:Kepsm=a}, $K^\epsilon_a(k,p)$ is
dominated by $\epsilon^{\gamma - \beta/2 - 1} k^{1+\alpha}
p^{2-\alpha}$ in $D_2$.  From~\eqref{eq:prop0}, 
$p^{2-\alpha}c_{a+1}^*(p)c_a(k)$ is integrable over the region $k,p\ge 0$.
Therefore, the contribution from $D_2$ to~\eqref{eq:3parts} 
is dominated by
$\epsilon^{2\gamma - \case12\beta - 1}$ (the additional factor of $\epsilon^\gamma$ is due to the fact that
the integral is confined to $|k^2 - p^2| \le \epsilon^\gamma)$. Given the choice of $\gamma$, this
vanishes as $\epsilon\rightarrow 0$.

The remaining region  $D_3$ is given by $k,p\ge 0$, $k^2 + p^2 \ge \epsilon^\beta$ and
$|p^2 - k^2| \ge \epsilon^\gamma$.  Integrating by parts with respect to
$\tau$ in (\ref{eq:Kepsm=a}), we get that
\begin{eqnarray}
  \label{eq:byparts}
\fl K^\epsilon_a(k,p) = \frac{1}{\pi} k^{1+\fracalph}
p^{2-\fracalph} (p^2-k^2) \nonumber\\
\lo \times \re \left(  \left.\frac{\rme^{-S/\epsilon^2}}{S_\tau}\right|_{\tau = 0} -
 \left.\frac{\rme^{-S/\epsilon^2}}{S_\tau}\right|_{\tau=\pi/2} -
  \ \int_0^{\pi/2}  \rme^{-S} \frac{S_{\tau\tau}}{S_{\tau}^2}\,\rmd \tau\right).
\end{eqnarray}
The first term gives
\begin{equation}
  \label{eq:firstterm}
  \frac{2}{\pi} \sin\pi\fracalph\, k^{1+\fracalph}p^{2-\fracalph} \left(1 + \frac{4(1-\fracalph)\epsilon^2}{p^2-k^2}
\right)^{-1}.
\end{equation}
Its contribution to the integral over $D_3$ in~\eqref{eq:3parts} yields, in the limit $\epsilon\rightarrow 0$,
the required result~\eqref{eq:mea}.

It remains to show that the contribution to the $D_3$-integral from the
remaining terms in~\eqref{eq:byparts} vanishes in the limit $\epsilon\rightarrow0$,
It is readily seen that the contribution from the second term vanishes
exponentially with $\epsilon$. For the third term, we note that, on
the interval $0\le \tau\le \pi/2$, $S_{\tau\tau}/S_{\tau}^2$ is
dominated by $(k^2 + p^2)/(k^2 - p^2)^2$, and $\exp(-S/\epsilon^2)$ is
dominated by $\exp(-\epsilon^{\beta-2}\tau^2/2)$.  Therefore, the
integral $\int_0^{\pi/2} \rme^{-S/\epsilon^2}
(S_{\tau\tau}/S_\tau^2)\rmd \tau$ is dominated by
$\epsilon^{1-\beta/2}(k^2 + p^2)/(p^2-k^2)^2$. Thus, the third term
in~\eqref{eq:byparts} is dominated by $\epsilon^{1-\beta/2}(k^2 +
p^2)k^{1+\fracalph}p^{2-\fracalph}/|k^2-p^2|$, which on $D_3$ is
dominated by $\epsilon^{1-\beta/2-\gamma}(k^2 + p^2)^{7/2}$.
The contribution to the integral over $D_3$ in~\eqref{eq:3parts} is
dominated by $\epsilon^{1-\beta/2-\gamma}$, which, by the choice
of $\gamma$, vanishes as $\epsilon\rightarrow 0$.

\References
\item Abramowitz M and Stegun IA 1970 {\it Handbook of Mathematical
    Functions} (New York: Dover).
\item Aharonov Y and Bohm D 1959 \PR  {\bf 115} 485.
\item Berry MV 1999 \JPA {\bf 32} 5627.
\item Gradshteyn IS and Ryzhik IM 1980 {\it Tables of Integrals,
    Series and Products}, (London: Academic Press).
\item Iordanskii SV 1966 {\it Sov. Phys. JETP} {\bf 22} 160.
\item Landau LD and Lifshitz EM 1965 {\it Quantum Mechanics} (Oxford:
  Pergamon Press).
\item Nielsen M and Hedeg\aa ard P 1995 \PR B {\bf 51} 7679--7699.
\item Olariu S and Popescu II 1983 \PR D {\bf 27} 383 --94.
\item Olariu S and Popescu II 1985 \RMP {\bf 57} 339 --436.
\item Peshkin M 1986 {\it Phys.~Rep.} {\bf 80} 375 -- 86.
\item Peshkin M 1989, in Peshkin M and Tonomura A 1989 {\it The Aharonov-Bohm Effect}
  (Berlin: Springer-Verlag).
\item Shelankov A 1998 {\it Europhys.~Lett.} {\bf 43}, 623 -- 8.
\item Shelankov A 2000, preprint.
\item Sonin EB 1976 {\it Sov. Phys. JETP} {\bf 42}, 469.
\item Sonin EB 1997 {\it Phys. Rev.} B {\bf 55}, 485.
\item Stone M 2000 \PR B {\bf 61} 11780 -- 6.
\item Thouless DJ, Ao P, Niu Q, Geller MR, Wexler C, {\it The 9th International Conference
on Recent Progress in Many-Body Theories, 21 -- 25 July 1997, Sydney, Australia} (cond-mat/9709127).

\endrefs
\end{document}